\documentclass[aps,pra,twocolumn,showpacs,preprintnumbers,amsmath,amssymb,superscriptaddress,floatfix]{revtex4-1}

\usepackage{amssymb}
\usepackage{graphicx}
\usepackage{dcolumn}
\usepackage{bm}
\usepackage{psfrag}
\usepackage{color}
\usepackage{amsmath}
\usepackage{amsthm}
\usepackage{hyperref}
\usepackage{booktabs}
\usepackage{makecell}

\makeatletter
\def\subsubsection{\@startsection{subsubsection}{3}{10pt}{-1.25ex plus -1ex minus -.1ex}{0ex plus 0ex}{\normalsize\bf}}
\def\paragraph{\@startsection{paragraph}{4}{10pt}{-1.25ex plus -1ex minus -.1ex}{0ex plus 0ex}{\normalsize\textit}}
\renewcommand\@biblabel[1]{#1}
\renewcommand\@makefntext[1]%
{\noindent\makebox[0pt][r]{\@thefnmark\,}#1}
\DeclareRobustCommand\onlinecite{\@onlinecite}
\def\@onlinecite#1{\begingroup\let\@cite\NAT@citenum\citealp{#1}\endgroup}
\def\tagform@#1{\maketag@@@{\ignorespaces#1\unskip\@@italiccorr}}
\let\orgtheequation\theequation
\def\theequation{(\orgtheequation)}
\makeatother

\begin{document}

\title{Universal time-evolution of a Rydberg lattice gas with perfect blockade}

\author{B. Olmos}
\email{beatriz.olmos-sanchez@nottingham.ac.uk}
\affiliation{Midlands Ultracold Atom Research Centre - MUARC, The University of Nottingham, School of Physics and Astronomy, Nottingham, United Kingdom}
\author{R. Gonz\'{a}lez-F\'{e}rez}
\email{rogonzal@ugr.es}
\affiliation{Instituto `Carlos I' de F\'{\i}sica Te\'orica y Computacional and Departamento de F\'{\i}sica At\'omica, Molecular y Nuclear, Universidad de Granada, 18071 Granada, Spain}
\author{I. Lesanovsky}
\email{igor.lesanovsky@nottingham.ac.uk}
\affiliation{Midlands Ultracold Atom Research Centre - MUARC, The University of Nottingham, School of Physics and Astronomy, Nottingham, United Kingdom}
\author{L. Vel\'{a}zquez}
\email{velazque@unizar.es}
\affiliation{Departamento de Matem\'{a}tica Aplicada, CPS, Universidad de Zaragoza, 50018 Zaragoza, Spain}

\date{\today}

\begin{abstract}
We investigate the dynamics of a strongly interacting spin system that is motivated by current experimental realizations of strongly interacting Rydberg gases in lattices. In particular we are interested in the temporal evolution of quantities such as the density of Rydberg atoms and density-density correlations when the system is initialized in a fully polarized state without Rydberg excitations. We show that in the thermodynamic limit the expectation values of these observables converge at least logarithmically to universal functions and outline a method to obtain these functions. We prove that a finite one-dimensional system follows this universal behavior up to a given time. The length of this universal time period depends on the actual system size. This shows that already the study of small systems allows to make precise predictions about the thermodynamic limit provided that the observation time is sufficiently short. We discuss this for various observables and for systems with different dimensions, interaction ranges and boundary conditions.
\end{abstract}

\maketitle

\section{Introduction}
Current experimental and theoretical progress in the control of ultra cold gases of Rydberg atoms has revealed remarkable insights into the dynamic and static properties of strongly interacting many-body systems \cite{Gallagher84,Saffman10-2}. The strong interaction between atoms in highly excited (Rydberg) states can be ten orders of magnitude larger than the interaction between ground state atoms and thus make ensembles of Rydberg atoms particularly suited for the study of strongly correlated phenomena. The emerging strong correlations are most strikingly reflected in the so-called Rydberg blockade \cite{Jaksch00,Lukin01} - a pronounced suppression of the probability to excite a Rydberg atom in the vicinity of another. A number of experiments have succeeded in demonstrating this excitation blockade and provided evidence for the coherent nature of the excitation dynamics in an impressive fashion \cite{Tong04,Singer04,Vogt06,Raitzsch08,Urban09,Gaetan09}.
From the theoretical side, ensembles of Rydberg atoms can often be accurately modeled by spin Hamiltonians. This approach has proven to be an extremely valuable tool and has delivered insights into the correlated ground state phases \cite{Pohl10,Schachenmayer10,Weimer10-2,Mukherjee11,Lesanovsky11,Ji11,Lesanovsky12,Zeller12} as well as into the dynamics of these systems \cite{Weimer08,Sun08,Olmos09-3,Olmos10-1,Giscard11,Mayle11,Ates11-2,Heeg12}.
\begin{figure}[h!!]
\includegraphics[width=\columnwidth]{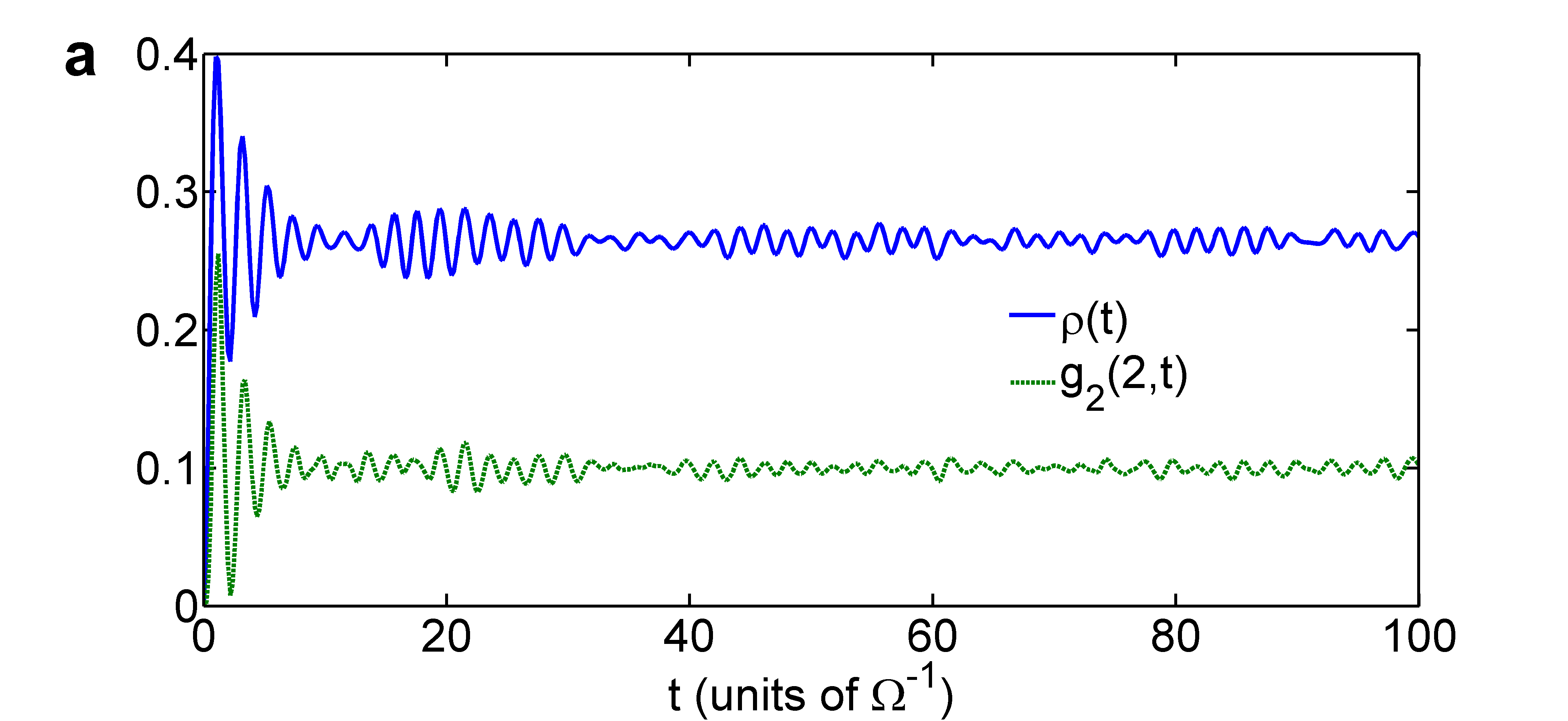}
\includegraphics[width=\columnwidth]{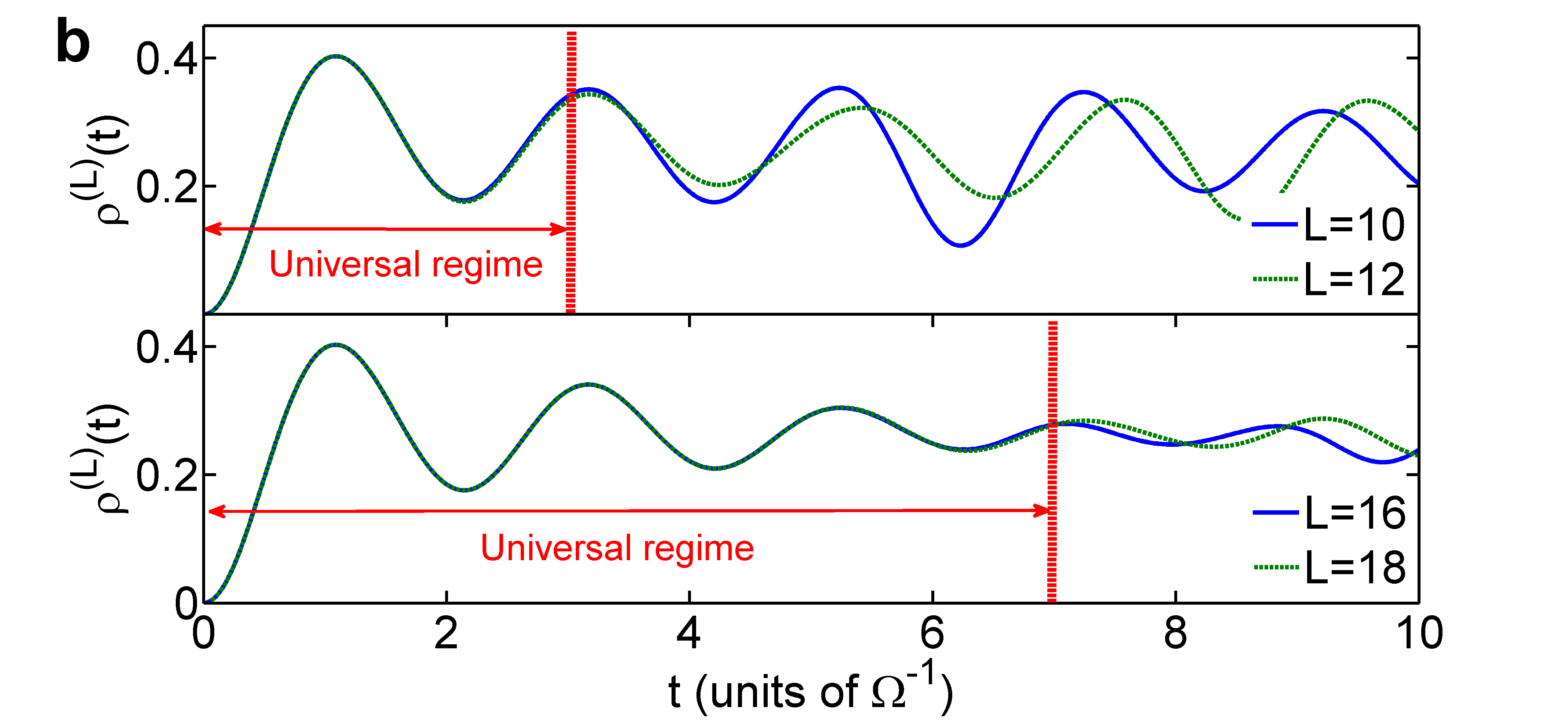}
\caption{(Color online) \textbf{a}: General behavior of the time evolution of observables such as the density of excited atoms $\rho(t)$ or the density-density correlation functions $g_2(d=2,t)$ between atoms separated by two sites (lattice with $L=25$ sites). \textbf{b}: Short time behavior of the density of excited atoms. In the upper panel, for lattice sizes $L=10$ and $12$. In the lower one same quantity for $L=16$ and $18$. One can observe how the universal regime grows with $L$: it is given by $t \lesssim 3$ for $L \gtrsim 10$, and by $t \lesssim 7$ for $L \gtrsim 16$.}
\label{Fig:universalregime}
\end{figure}

A particularly simple spin-model describing the quantum dynamics of an arrangement of Rydberg atoms was explored in Refs. \cite{Olmos09-1,Olmos09-2,Ates11-2} by some of the authors. There we considered atoms placed on a one-dimensional regular lattice - a situation which since very recently can also be achieved experimentally \cite{Viteau11,Anderson11}. We moreover assumed a perfect excitation blockade for neighboring sites, i.e., no two neighboring atoms can be simultaneously excited to the Rydberg state. For this simplified situation, we studied in particular the time-evolution of local observables such as the density of Rydberg atoms and density-density correlation functions. We found that the temporal behavior of all of these observables is qualitatively similar: The short time dynamics is characterized by strong oscillations while for long times a steady state is established in which all quantities fluctuate with small amplitude around a constant value (see Fig. \ref{Fig:universalregime}a). The focus of these previous works was on the long-time behavior, i.e., on the investigation of the origin of this (seemingly thermal) steady state. Indeed, in Refs. \cite{Ates11-2,Ates12} we showed that this steady state can be understood as a maximum entropy state and that the evolution into it is governed by a Fokker-Planck equation.

The aim of this work is to study the short-time behavior of observables in a Rydberg lattice gas. In particular, we will answer the question of how well the numerical results obtained for finite systems approximate the corresponding results in the thermodynamic limit. We find that for sufficiently short times expectation values of physical quantities are independent of the system size and, hence, universal. An example for that is depicted in Fig. \ref{Fig:universalregime}b with the evolution of the density of Rydberg atoms. The universal time interval grows with increasing system size. We show that there are universal functions for the expectation values of general observables in the thermodynamic limit to which the results of finite systems converge at least logarithmically. This means that even the study of finite systems gives access to properties of the thermodynamic limit provided that the propagation time is sufficiently short. We discuss which observables exhibit this convergence to the universal behavior, and the validity of these results for systems with different geometries, interaction ranges and dimensions.
Note that, based on the so-called Lieb-Robinson bounds \cite{Lieb72}, the existence of a thermodynamic limit was mathematically proven for quantum lattice systems with polynomially decaying interactions \cite{Nachtergaele06,Eisert06,Bravyi06}. In this work, we calculate the rate of convergence to this thermodynamic limit and show that it is, at least, logarithmic. We provide a systematic procedure to obtain these universal functions using the particularly symmetric case of a ring lattice.

The paper is structured as follows. In Section \ref{Sec:H}, we describe in detail the system under study and present a derivation of the Hamiltonian that governs its dynamics. In Section \ref{Sec:general}, we discuss the properties of the Hamiltonian and their implications in the time-evolution of the expectation value of any general observable. In addition, we introduce some of the mathematical tools we will use throughout the remainder of the paper. In Section \ref{Sec:ring}, we focus on the case of a one-dimensional lattice with periodic boundary conditions and any blockade radius. Here, we prove that the time-evolution of the expectation value of a set of operators (including the Rydberg density) in a finite size system converges at least logarithmically for all times to a universal function (i.e., independent of the size of the system) in the thermodynamic limit. The technique to derive this universal function is also described here. In Section \ref{Sec:seg}, we analyze the same system but with open boundary conditions and show that the results found in the periodic case are equally valid. Finally, we discuss the generalization of these results to higher dimensions in Section \ref{Sec:dim}. The conclusions and open problems are given in Section \ref{Sec:con}. In Appendices \ref{App:univ}, \ref{App:words} and \ref{App:kappa}, we provide the bulk of the mathematical proof of the existence of the universal functions and the convergence to those of the finite-size expectation values. In Appendix \ref{App:matrixsegment}, we present the method to create the matrix representation of some operators on the one-dimensional linear lattice.

\section{The system and its Hamiltonian} \label{Sec:H}
The system we are considering is a one-dimensional lattice with $L$ sites and spacing $a$. Each site shall contain a single atom which occupies the (local) vibrational ground state. The ground state $\left|g\right>$ and a highly excited (Rydberg) state $\left|r\right>$ of each atom (considered as two-level systems) are coupled resonantly by means of a laser with Rabi frequency $\Omega$. The atoms in the Rydberg state interact via the van-der-Waals potential $C_6/\left|\mathbf{r}\right|^6$, where $C_6$ is the van-der-Waals coefficient \cite{Singer05} and $|\mathbf{r}|$ is the distance between them. Having only two possible states reduces the problem to a spin-1/2 system where one can identify $\left|g\right>=r\left|r\right>\equiv\left| \downarrow\right>$ and $\left|r\right>=r^\dagger\left|g\right>\equiv\left| \uparrow\right>$. The annihilation and creation operators $r_k$, $r_k^\dag$ of a Rydberg state at site $k$ satisfy thus a spin-$1/2$ algebra
\begin{eqnarray*}
  &&[r_j,r_k]=[r_j^\dag,r_k^\dag]=[r_j,r_k^\dag]=0,\quad j\neq k, \\
  &&r_k^2={r_k^\dag}^2=0, \quad \{r_k,r_k^\dag\}=1.
\end{eqnarray*}
With these operators the Hamiltonian of the system reads
\begin{equation*}
  {\cal H} =\Omega\sum_{k=1}^L(r_k+r_k^\dagger)+V\sum_{k\neq j}\frac{n_kn_j}{\left|k-j\right|^6},
\end{equation*}
where $n_k = r_k^\dag r_k$ counts the number of Rydberg atoms at site $k$ and $V=C_6/(2a^6)$ represents the nearest neighbor interaction strength divided by two to compensate the double counting in the sum.

\begin{figure}
\includegraphics[width=0.9\columnwidth]{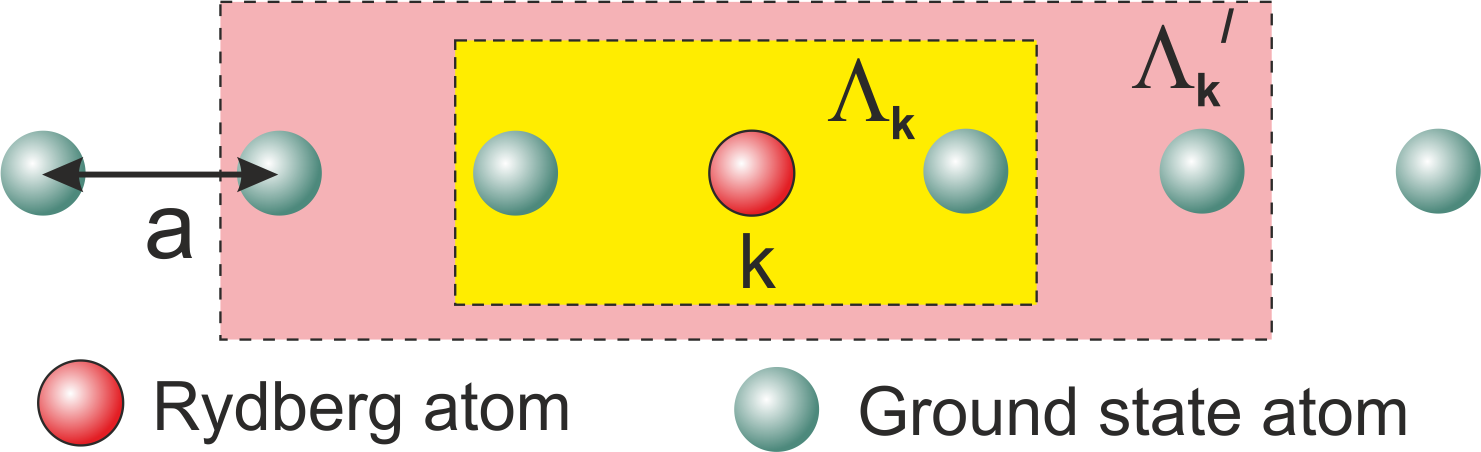}
\caption{(Color online) Perfect blockade model. We consider a cutoff of the interaction such that only the atoms inside a domain $\Lambda_k$ interact with an atom excited on the $k$-th site. Two different examples of domains are shown: nearest neighbors ($\Lambda_k$) and next-nearest neighbors ($\Lambda'_k$) interaction. Moreover, we consider the interaction between excited states so strong that within the corresponding domain only one atom can be excited to the Rydberg state.}
\label{Fig:lambda}
\end{figure}
In the following we will perform two approximations which lead us to the perfect blockade model. First, we do not consider the full $\left|\mathbf{r}\right|^{-6}$-tail of the van-der-Waals interaction but rather a cut-off at a radius $R_b=\lambda_ba$, i.e., only atoms on sites contained in the subset $\Lambda_k=\left\{j:|k-j|\leq \lambda_b\;\mathrm{and}\;j\neq k\right\}$ interact with an excited atom located at the $k$-th site (see Fig. \ref{Fig:lambda}). The Hamiltonian can then be rewritten as ${\cal H}=H_0+H_\mathrm{int}$ with
\begin{equation*}
  H_0=\Omega\sum_{k=1}^L(r_k+r_k^\dagger), \quad H_\mathrm{int}=\sum_{k=1}^L\sum_{j\in\Lambda_k}V_{kj} n_kn_j,
\end{equation*}
where $V_{kj}=V/|k-j|^6$. In addition, we assume that the interaction energy between atoms inside the radius $R_b$ - often referred to as the blockade radius - is much larger than the Rabi frequency of the laser, i.e., $\Omega\ll V/\lambda_b^6$. This means in turn that the probability of having two atoms simultaneously excited within the set $\Lambda_k$ is strongly suppressed. These two assumptions can only be satisfied if
\begin{equation} \label{Eqn:b-cond}
  \frac{V}{\lambda_b^6}\gg\Omega\gg\frac{V}{(\lambda_b+1)^6}.
\end{equation}
Note that these conditions are more difficult to meet the larger the number of sites inside the \emph{blockade} radius $R_b$.

In the next step, we make the excitation blockade manifest by transforming the Hamiltonian into the interaction picture with the unitary operator $U=e^{-iH_\mathrm{int}t}$ \cite{Lesanovsky11}. This results in a new Schr\"odinger equation with effective Hamiltonian
\begin{equation*}
  H=\Omega\sum_{k=1}^Le^{2itn_k\sum_{j\in\Lambda_k}V_{kj}n_j} (r_k+r_k^\dag) \, e^{-2itn_k\sum_{j\in\Lambda_k}V_{kj}n_j},
\end{equation*}
where we have used that $V_{kj}=V_{jk}$. Since $k\notin\Lambda_k$, all the terms contained in the sum over $j\in\Lambda_k$ commute with $n_k$. Now, we have to calculate explicitly
\begin{equation*}
  e^{i\alpha_k n_k}(r_k+r_k^\dag)\,e^{-i\alpha_k n_k}=e^{i\alpha_k}r_k^\dag+e^{-i\alpha_k}r_k,
\end{equation*}
with $\alpha_k=2t\sum_{j\in\Lambda_k}V_{kj}n_j$. Note that $n_k^2=n_k$, i.e., $n_k$ and $m_k=1-n_k$ are projectors onto the subspaces with a Rydberg and a ground state atom on site $k$, respectively. Thus, one has
\begin{eqnarray*}
  e^{i\alpha_k}=\prod_{j\in\Lambda_k}e^{2itV_{kj}n_j}= \prod_{j\in\Lambda_k}\left(m_j+n_je^{2itV_{kj}}\right).
\end{eqnarray*}
Next, we neglect in this expression the terms that oscillate with frequencies $V_{kj}$ for all $j\in\Lambda_k$ which under our initial assumption are much larger than the Rabi frequency. This is the so-called rotating wave or secular approximation, and gives rise to the effective Hamiltonian
\begin{equation} \label{Eqn:Heff}
  H=\Omega\sum_{k=1}^L(r_k+r_k^\dag)M_k,
\end{equation}
with $M_k=\prod_{j\in\Lambda_k}m_j$. This Hamiltonian couples states differing in a single Rydberg excitation, but the factor $M_k$ prevents the creation of a Rydberg excitation at site $k$ whenever there is another already present at some site $j\in\Lambda_k$.

The natural initial state of the system is the one with all atoms in the ground state, i.e., $|\mathbf{0}\rangle=\bigotimes_k \left|g\right>_k$ and $M_k|\mathbf{0}\rangle=|\mathbf{0}\rangle$. The Hamiltonian then only couples $\left|\mathbf{0}\right>$ to states which have eigenvalue $1$ with respect to all $M_k$. As a consequence, only states with at most one excitation inside the neighborhoods $\left\{\Lambda_k\right\}_{k=1}^L$ are accessible by a time-evolution under $H$.

The generalization of the previous arguments to higher dimensional lattices leads to a Hamiltonian of the same form as \autoref{Eqn:Heff}. The higher dimensional version of condition \ref{Eqn:b-cond} is, however, more difficult to satisfy. For instance, in the case of a blockade radius $R_b=a$, \autoref{Eqn:b-cond} reads $V\gg\Omega\gg V/2^6$, while for a 2-dimensional square and a 3-dimensional cubic lattices, it becomes the more stringent condition $V\gg\Omega\gg V/2^3$. We would like to remark that even when the conditions for the validity of the effective Hamiltonian \ref{Eqn:Heff} are not strictly satisfied, such a Hamiltonian provides a first approximation to the dynamics of a Rydberg lattice gas which can be corrected by perturbation techniques.

In the next sections, we will mathematically  prove that the dynamics of a system driven by the Hamiltonian \ref{Eqn:Heff} exhibits some universal features, which do not depend on the lattice geometry or the interaction range.

\section{Time-evolution of observables: general considerations} \label{Sec:general}
Our analysis begins by discussing a property of the system that holds regardless of the geometry of the lattice: The operator $(-1)^n$, with $n=\sum_kn_k$ being the total number of atoms in the Rydberg state, satisfies
\begin{equation*}
  \{(-1)^{n_k},r_k\}=\{(-1)^{n_k},r_k^\dag\}=0
\end{equation*}
which in turn yields
\begin{equation}\label{Eqn:anticomm}
  \{(-1)^n,H\}=0.
\end{equation}
This anticommutation relation implies that $H$ only couples states whose total number of Rydberg excitations have opposite parity $(-1)^n$. Although $(-1)^n$ is not a symmetry of $H$, the fact that both operators anticommute has some implications in the corresponding spectrum and dynamics.

First, if $|\Psi\rangle$ is an eigenvector of $H$ with eigenvalue $E$, the state $(-1)^n|\Psi\rangle$ is also an eigenvector, but with eigenvalue $-E$. Therefore, the energy spectrum of $H$ is symmetric with respect to $E=0$. Moreover, any eigenvector $|\Psi\rangle$ with eigenvalue $E\neq0$ must be orthogonal to $(-1)^n|\Psi\rangle$. Thus, $\langle\Psi|(-1)^n|\Psi\rangle=0$, which yields $\langle\Psi_e|\Psi_e\rangle=\langle\Psi_o|\Psi_o\rangle=1/2$ where $|\Psi_e\rangle$ and $|\Psi_o\rangle$ are the projections of $|\Psi\rangle$ onto the subspaces with even and odd total number of Rydberg states, respectively. We conclude from this that any eigenvector of $H$ with non zero eigenvalue has equal probability of having an odd or even number of Rydberg excitations.

For some particular choices of the lattice geometry and the blockade range $\Lambda_k$, the system and its Hamiltonian $H$ possess additional symmetries. For instance, in the case of a ring lattice (one-dimensional lattice with periodic boundary conditions) the system is invariant under translations and reversal of the sites. Any symmetry of the system splits the state space into several $H$-invariant subspaces (the eigenspaces of the symmetry), decoupling the evolution into lower dimensional ones with the same spectral properties described previously for $H$. If this symmetry also commutes with $(-1)^n$, the corresponding subspaces are $(-1)^n$-invariant too. As a consequence, any odd-dimensional subspace which is simultaneously invariant for $H$ and $(-1)^n$ must enclose a zero energy eigenstate of $H$. Otherwise, the number of orthogonal eigenvectors of the reduced Hamiltonian should be even due to the symmetry of its spectrum with respect to the origin.

The anticommutation relation \ref{Eqn:anticomm} also bears implications for the dynamics. In the Heisenberg picture the evolution of any observable $A$ is given by $A(t)=e^{itH}A\,e^{-itH}$. The evolution of this arbitrary observable $A$ is an entire function of $t$, i.e., an analytic function on the whole complex plane when $t$ is considered as a complex variable, so that its power series must converge for any value of $t$. The corresponding power series expansion yields
\begin{equation}\label{Eqn:seriesA}
  A(t)=\sum_{j=0}^\infty \frac{1}{j!} \, i^j t^j \mathrm{ad}_H^j(A),\\
\end{equation}
where we have defined the adjoint endomorphism
\begin{equation*}
  \mathrm{ad}^j_H(A)\equiv[H,A]_j=\left[H,[H,A]_{j-1}\right], \qquad\mathrm{ad}^0_H(A)\equiv A.
\end{equation*}
Let us assume that $[A,(-1)^n]=0$, i.e., the operator $A$ conserves the parity of the number of excitations. Then, the relation \ref{Eqn:anticomm} yields
\begin{equation*}
  (-1)^n A(t) (-1)^n = A(-t).
\end{equation*}
If $|\Phi\rangle$ and $|\Psi\rangle$ are two eigenstates of $n$ with total number of Rydberg states $n_\Phi$ and $n_\Psi$, respectively, the matrix element $\langle \Phi | A(t) | \Psi \rangle$ satisfies
\begin{equation*}
\langle \Phi | A(-t) | \Psi \rangle = (-1)^{n_\Phi+n_\Psi} \langle \Phi | A(t) | \Psi \rangle.
\end{equation*}
As a consequence, $\langle \Phi | A(t) | \Psi \rangle$ is an even or an odd function of $t$ depending whether $n_\Phi$ and  $n_\Psi$ have the same parity or not. Setting $|\Phi\rangle=|\Psi\rangle$, we find that the expectation value of $A$ starting from an initial state with a definite number of Rydberg excitations is an even function of $t$.

We now apply these results to the time-evolution of the expectation value of the total number of Rydberg atoms $n$. We consider the initial state $\left|\mathbf{0}\right>$, i.e., $\langle n(t) \rangle \equiv \langle \mathbf{0} | n(t) | \mathbf{0} \rangle$, so that $\langle n(0) \rangle=0$. By using \autoref{Eqn:seriesA}, we see that only the even terms contribute to the expansion:
\begin{equation}\label{Eqn:seriesn}
  \langle n(t) \rangle = \sum_{j=1}^\infty c_j \, t^{2j}, \quad c_j = \frac{(-1)^j}{(2j)!} \, \langle\mathrm{ad}_{H}^{2j}\!(n)\rangle.
\end{equation}
The first term of the sum is proportional to $t^2$ and can be calculated using the algebra of $r_k$ and $r_k^\dag$ as
\begin{eqnarray*}
  &&\mathrm{ad}_{H}^{1}\!(n)=[H,n]=\Omega \sum_k (r_k-r_k^\dag) M_k\\
  &&\mathrm{ad}_{H}^{2}\!(n)=[H,[H,n]]= 2\Omega^2\sum_k (n_k-m_k) M_k\\
  && \kern30pt + 2 \Omega^2 \sum_k \sum_{j\in\Lambda_k} (r_j r_k^\dag - r_j^\dag r_k)  M_j M_{k \setminus j}
\end{eqnarray*}
with $M_{k \setminus j} = \prod_{l\in\Lambda_k,\, \scriptstyle l\neq j } m_l$. Taking the expectation value in the last expression we obtain $\langle\mathrm{ad}_{H}^{2}\!(n)\rangle =-2 \Omega^2 L$ and in the expansion \ref{Eqn:seriesn} the first coefficient is $c_1=\Omega^2L$. Therefore, the expectation value of the Rydberg density defined as $\rho(t)=\langle n(t)\rangle/L$ is an even function of $t$ with the universal behavior $\rho(t)=\Omega^2 t^2 + \mathcal{O}(\Omega^4 t^4)$, regardless of the dimension or the geometry of the lattice, and for any blockade range.

\emph{In the remainder of the paper we will search for such a universality (independence of the size of the system) in  higher terms of the series expansion of $\rho(t)$, as well as for expectation values of other operators.} To tackle this task, we will extensively use the following property: If $A$ and $B$ are self-adjoint, $\mathrm{ad}_{B}^j(A)$ is self-adjoint or anti self-adjoint depending on whether $j$ is even or odd. Consequently, to obtain $\mathrm{ad}_B^j(A)=[B,\mathrm{ad}_B^{j-1}(A)]$ we can use the identities
\begin{equation*}
  \mathrm{ad}_B^j(A)=\left\{\begin{array}{lr}
    B \, \mathrm{ad}_B^{j-1}(A) + \mathrm{h.c.} & \qquad j\,\mathrm{even}\\
    B \, \mathrm{ad}_B^{j-1}(A) - \mathrm{h.c.} & \qquad j\,\mathrm{odd},
  \end{array}\right.
\end{equation*}
i.e., the symmetrization or anti-symmetrization of $B\,\mathrm{ad}_B^{j-1}(A)$ for even and odd $j$, respectively.

For convenience, we will set from now on the energy scale $\Omega=1$. The change $t \to \Omega t$ allows us to recover the dependence on $\Omega$ in the evolution.

\section{The one-dimensional ring lattice} \label{Sec:ring}
In this section, we will analyze the dynamics for a ring lattice with only nearest neighbor interaction, that is, a one-dimensional finite lattice with periodic boundary conditions and blockade range $\Lambda_k=\{k-1,k+1\}$. This system was previously explored in Refs. \cite{Sun08,Olmos09-1,Olmos09-2,Ates11}.

The physical properties of this system will in general depend on the number of sites $L$. We make apparent this explicit dependence on $L$ by including a superscript $(L)$. The Hamiltonian reads
\begin{equation}\label{Eqn:Hring}
  H^{(L)}=\sum_{k=1}^L H_k,\qquad\mathrm{with}\qquad H_k= m_{k-1}\Delta_k\,m_{k+1}.
\end{equation}
where the notation $\Delta_k = r_k + r_k^\dag$ has been introduced. According to the discussion in \autoref{Sec:general}, the Rydberg density is given by
\begin{equation}\label{Eqn:rho_L}
  \rho^{(L)}(t)= \frac{1}{L}\langle n^{(L)}(t)\rangle=\sum_{j=1}^\infty c_j^{(L)} t^{2j},
\end{equation}
with $n^{(L)} = \sum_{k=1}^L n_k$, and
\begin{equation} \label{Eqn:cj}
c_j^{(L)} = \frac{(-1)^j}{(2j)!} \, \langle \mathrm{ad}_{H^{(L)}}^{2j}(n_k) \rangle  \qquad\forall k,
\end{equation}
where the invariance of the system under translations has been used. The periodic boundary conditions imply that any index $k$ must be understood mod($L$).

In the previous Section, it has been already shown that in general $c_1^{(L)}=1$. We calculate again this coefficient using a procedure that allows us to obtain $c_j^{(L)}$ beyond $j=1$. First, we need to compute
\begin{equation*}
  \mathrm{ad}_{H^{(L)}}^1(n_k) = H_k n_k - \mathrm{h.c.},
\end{equation*}
where we have used that $H_k$ is the only term of $H^{(L)}$ which does not commute with $n_k$. This anti-symmetrization yields
\begin{equation} \label{Eqn:ad1}
\mathrm{ad}_{H^{(L)}}^1(n_k) = m_{k-1} (r_k-r_k^\dag) m_{k+1}.
\end{equation}
In order to compute $\mathrm{ad}_{H^{(L)}}^2(n_k)$, we first realize that among the terms of $H^{(L)}$ only $H_{k-1}$, $H_k$ and $H_{k+1}$ do not commute with $\mathrm{ad}_{H^{(L)}}^1(n_k)$, and hence
\begin{equation*}
  \mathrm{ad}_{H^{(L)}}^2(n_k)=\left(\sum_{j=-1}^1 H_{k+j}\right) \, m_{k-1} (r_k-r_k^\dag) m_{k+1} + \mathrm{h.c.}.
\end{equation*}
Upon symmetrization, we obtain
\begin{equation} \label{Eqn:ad2}
\begin{aligned}
& \mathrm{ad}_{H^{(L)}}^2(n_k) = 2a_k^{(1)} + a_k^{(2)} + a_{k+1}^{(2)},
\\
& a_k^{(1)} = m_{k-1} \delta_k m_{k+1},
\\
& a_k^{(2)} = m_{k-2} (r_{k-1} r_k^\dag + r_{k-1}^\dag r_k) m_{k+1},
\end{aligned}
\end{equation}
with  $\delta_k=n_k-m_k$. Hence, $\langle \mathrm{ad}_{H^{(L)}}^2(n_k) \rangle = 2\langle \delta_k \rangle = -2$ and we recover the expected result $c_1^{(L)}=1$.

We now proceed with the second coefficient $c_2^{(L)}$. For  $\mathrm{ad}_{H^{(L)}}^3(n_k)$, the same procedure as before yields
\begin{equation*}
  \mathrm{ad}_{H^{(L)}}^3(n_k) = 2A_k^{(1)} + A_k^{(2)} + A_{k+1}^{(2)},
\end{equation*}
where
\begin{equation*}
  A_k^{(1)} = [\sum_{j=-1}^1 H_{k+j},a_k^{(1)}], \quad A_k^{(2)} = [ \sum_{j=-2}^1 H_{k+j},a_k^{(2)}].
\end{equation*}
Calculating the commutators $A_k^{(1)}$ and $A_k^{(2)}$ and performing the anti-symmetrization finally results in
\begin{equation}\label{Eqn:ad3}
\begin{aligned}
& \mathrm{ad}_{H^{(L)}}^3(n_k) = 4b_k^{(1)} + b_k^{(2)} + 3b_{k+1}^{(2)} + 3b_k^{(3)} + b_{k+1}^{(3)}
\\
& \kern58pt + b_k^{(4)} + 2b_{k+1}^{(4)} + b_{k+2}^{(4)},
\\
& b_k^{(1)} = m_{k-1} (r_k-r_k^\dag) m_{k+1}
\\
& b_k^{(2)} = m_{k-2} m_{k-1} (r_k-r_k^\dag) m_{k+1},
\\
& b_k^{(3)} = m_{k-2} (r_{k-1}-r_{k-1}^\dag) m_k m_{k+1},
\\
& b_k^{(4)} = m_{k-3} (r_{k-2}^\dag r_{k-1} r_k^\dag - r_{k-2} r_{k-1}^\dag r_k) m_{k+1}.
\end{aligned}
\end{equation}
As we can see, the level of complexity of the terms $\mathrm{ad}_{H^{(L)}}^j(n_k)$ increases very quickly. However, since we are only interested in $c^{(L)}_2$ it is not necessary to compute $\mathrm{ad}_{H^{(L)}}^4(n_k)$ completely. The only terms of $\mathrm{ad}_{H^{(L)}}^4(n_k)$ which contribute to $\langle\mathrm{ad}_{H^{(L)}}^4(n_k)\rangle$ are the products of operators $m_l$. Such terms are self-adjoint and, hence, they are generated simultaneously with equal coefficients in both summands of $H^{(L)}\mathrm{ad}_{H^{(L)}}^3(n_k) + \mathrm{h.c.}$.
\begin{table}
\caption{The most frequent operator multiplications $ab$ used through the text.\label{Eqn:mr}}
\begin{ruledtabular}
\begin{tabular}{c|cccc}
$a\backslash b$ & $r_j$ &$ r_j^\dag$ &$n_j$ & $m_j$ \\
\hline
$m_j$ &  $r_j$ & $0$ &$0$ &$m_j$ \\
$\Delta_j$ & $n_j$  & $ m_j$ & $r_j^\dag$& $r_j$ \\
\end{tabular}
\end{ruledtabular}
\end{table}
Using the algebra of $r_j$ and $r_j^\dag$ (see \autoref{Eqn:mr}) the terms of $H^{(L)}\mathrm{ad}_{H^{(L)}}^3(n_k)$ depending only on $m_l$ derive from the products of the form $H_kb_k^{(1)}$, $H_kb_k^{(2)}$ and $H_{k-1}b_k^{(3)}$. That is, the only part of $\mathrm{ad}_{H^{(L)}}^4(n_k)$ contributing to $c_2^{(L)}$ is the symmetrization of $4H_kb_k^{(1)} + H_kb_k^{(2)} + 3H_{k+1}b_{k+1}^{(2)} + 3H_{k-1}b_k^{(3)} + H_kb_{k+1}^{(3)}$. Hence, we find that $\langle \mathrm{ad}_{H^{(L)}}^4(n_k)\rangle = -24$ and, finally,
\begin{equation} \label{Eqn:c2L}
c_2^{(L)} = \frac{1}{4!} \langle\mathrm{ad}_{H^{(L)}}^4(n_k) \rangle = -1,
\end{equation}
also independent of the size of the lattice $L$.

\subsection{Universality condition of the expansion coefficients}
Neither $c_1^{(L)}$ nor $c_2^{(L)}$ have shown a dependence on the size of the ring $L$. This is, however, only true because we have implicitly assumed that $L>2$ and thus, the boundary effects due to the finite size of the system do not play a role.

Let us explore these finite size effects for the (artificial) case of $L=2$. The influence of the boundary becomes apparent in the quantity $\mathrm{ad}_{H^{(2)}}^j(n_1)$. While \autoref{Eqn:ad1} holds identifying $m_0=m_2$, i.e.,
\begin{equation*}
\mathrm{ad}_{H^{(2)}}^1(n_1)= [H_1,n_1]=m_0(r_1-r_1^\dag)m_2=(r_1-r_1^\dag)m_2,
\end{equation*}
the calculation of $\mathrm{ad}_{H^{(2)}}^2(n_1)$ gives
\begin{equation} \label{Eqn:ad2L=2}
\mathrm{ad}_{H^{(2)}}^2(n_1) = [H_1+H_2,m_0(r_1-r_1^\dag)m_2]
\end{equation}
and we see that \autoref{Eqn:ad2} is not valid due to two reasons: first, some contributions are missing if compared with the general expression $\mathrm{ad}_{H^{(L)}}^2(n_1)=[H_0+H_1+H_2,m_0(r_1-r_1^\dag)m_2]$; second, in the general case the commutators are calculated under the assumption that operators with different indices commute, but this only holds as long as there is no term with indices which are equal mod($L$), and this is violated by $m_0(r_1-r_1^\dag)m_2$ if $L=2$. If we proceed without simplifying $m_0$ and $m_2$, the calculation \ref{Eqn:ad2L=2} of $\mathrm{ad}_{H^{(2)}}^2(n_1)$ yields only part of the terms obtained in \autoref{Eqn:ad2} for the general case.

To calculate explicitly the coefficients $c_j^{(2)}$ for $L=2$, the corresponding $\mathrm{ad}_{H^{(2)}}^{2j}(n_1)$ are needed. For $j=1$, the result is
\begin{equation*}
 \mathrm{ad}_{H^{(2)}}^2(n_1) = 2 \delta_1 m_2 + r_1 r_2^\dag + r_1^\dag r_2,
\end{equation*}
so that
\begin{equation*}
c_1^{(2)} = - \frac{1}{2!} \langle\mathrm{ad}_{H^{(2)}}^2(n_1)\rangle = - \frac{1}{2!}(-2) = 1,
\end{equation*}
in agreement with the general arguments of \autoref{Sec:general}. However, for the next coefficient, we arrive at
\begin{eqnarray*}
\mathrm{ad}_{H^{(2)}}^3(n_1) &=& 5 (r_1-r_1^\dag) m_2 + 3 m_1 (r_2-r_2^\dag),
\\
\mathrm{ad}_{H^{(2)}}^4(n_1) &=& 10 \delta_1 m_2 + 6 m_1 \delta_2 + 4(r_1 r_2^\dag + r_1^\dag r_2),
\end{eqnarray*}
that yields a different result from \autoref{Eqn:c2L}, namely,
\begin{equation*}
c_2^{(2)} = \frac{1}{4!} \langle\mathrm{ad}_{H^{(2)}}^4(n_1)\rangle = \frac{1}{4!} (-16) = - \frac{2}{3}.
\end{equation*}
For $L=3$, Eqs. \ref{Eqn:ad1} and \ref{Eqn:ad2} remain valid because they involve commutators of $H^{(3)}=H_1+H_2+H_3$ and terms with no more than 3 consecutive indices, while \autoref{Eqn:ad3} fails. This only ensures that $c_1^{(3)}=1$. However, $c_2^{(3)}=-1$, coinciding with \autoref{Eqn:c2L}, because the perturbation created by the periodic boundary conditions does not affect the terms of $\mathrm{ad}_{H^{(3)}}^4(n_k)$ contributing to $c_2^{(3)}$.

The previous results seem to indicate that $\mathrm{ad}_{H^{(L)}}^j(n_k)$ and \emph{$c_j^{(L)}$ are universal for $j \leq L-1$} because they can be computed using only the universal algebra of $r_l$ and $r_l^\dag$ without being concerned with the boundary conditions. This is indeed true, and the proof is provided in Appendix \ref{App:univ}. We denote by $c_j$ the universal value of $c_j^{(L)}$ for $j \leq L-1$. For illustration purposes, the first five of these universal coefficients $c_j$ are given in \autoref{tb:coefficients}.
\begin{table}[h!!]
\caption{For blockade radius $R_b=a$ ($\lambda_b=1$), first five universal coefficients, $c_j$, of the Rydberg density $\rho(t)$, and of $\rho_d(t)$ with $d=2$ and $3$, $c_{d,j}$.
For a linear lattice, the first four universal coefficients $q_j$. For the blockade radii $\lambda_b=2$ and $3$, the first universal coefficients of the Rydberg density $\rho(t)$ are also displayed.}
\label{tb:coefficients}
\begin{center}
\begin{ruledtabular}
\begin{tabular}{@{}crrrrrr@{}}\toprule
& \multicolumn{4}{c}{$\lambda_b=1$}&$\lambda_b=2$&$\lambda_b=3$\\
\cline{2-5} \cline{6-6} \cline{7-7}
$j$ & $c_j$  &$c_{2,j}$ &$c_{3,j}$& $q_j$&$c_j$  &$c_j$  \\
\hline
$1$& $1$ &$0$&$0$&&$1$&$1$\\
$2$& $-1$ & $1$& $1$ &$\frac{2}{3}$&$-\frac{5}{3}$&$-\frac{7}{3}$\\
$3$& $\frac{3}{5}$& $-\frac{3}{2}$&$-2$&$\frac{38}{27}$&$\frac{77}{45}$&$\frac{152}{45}$\\
$4$& $-\frac{81}{280}$&$\frac{283}{240}$        &$\frac{61}{30}$&$\frac{518}{243}$&$-\frac{713}{504}$&\\
$5$& $\frac{3023}{25200}$&$-\frac{739}{1120}$&$-\frac{2393}{1680}$&$\frac{76016}{27207}$&&\\
\bottomrule
\end{tabular}
\end{ruledtabular}
\end{center}
\end{table}

\subsection{The universal Rydberg density $\rho(t)$} \label{SSub:universal}
At least formally, we can define now the power series
\begin{equation} \label{Eqn:rho}
\rho(t)=\sum_{j=0}^\infty c_j t^{2j}
\end{equation}
with $c_j$ being the universal expansion coefficients. We would naively expect that $\rho^{(L)}(t)\xrightarrow{L\to\infty}\rho(t)$, but this can not be simply  implied as a consequence of $c_j=\lim_{L\to\infty}c_j^{(L)}$. In fact, for the moment we cannot even be sure that the power series in \autoref{Eqn:rho} has a non null radius of convergence. Let us show here that \autoref{Eqn:rho} defines an entire function such that $\rho^{(L)}(t)\xrightarrow{L\to\infty}\rho(t)$ for any value of $t$. This implies that, although $\rho(t)$ is defined by a power series around the origin, it encodes the universality of $\rho^{(L)}(t)$ in any regime. We follow here a complementary approach to that presented in Ref. \cite{Nachtergaele06} that will also allow us to calculate the rate of convergence to $\rho(t)$, which will be referred to as the \emph{universal Rydberg density} in the remainder of the paper.

An analytic expression for all the coefficients $c_j$ and $c_j^{(L)}$ is not available. However, to prove the above statements it is enough to obtain bounds $|c_j|, |c_j^{(L)}| \leq b_j$ for a set of coefficients $b_j$ such that the power series $\sum_{j=1}^\infty b_jz^j$ has an infinite radius of convergence. Since  $\sum_{j=1}^\infty|c_j||t|^{2j} \leq \sum_{j=1}^\infty b_j|t|^{2j}$, then $\sum_{j=1}^\infty c_jt^{2j}$ is absolutely convergent for any $t$. Besides, bearing in mind that $c_j^{(L)}=c_j$ for $j \leq L-1$, we obtain
\begin{equation} \label{Eqn:conv}
|\rho(t)-\rho^{(L)}(t)| \leq \sum_{j=L}^\infty |c_j-c_j^{(L)}| |t|^{2j} \leq 2 \sum_{j=L}^\infty b_j |t|^{2j},
\end{equation}
which shows that $\rho^{(L)}(t)\xrightarrow{L\to\infty}\rho(t)$ at any time $t$. In Appendix \ref{App:words} we prove that $|c_j|,|c_j^{(L)}| \leq b_j$ for the coefficients
\begin{equation} \label{Eqn:b}
b_j = 2 \cdot 6^{2j-1} \frac{\kappa_{2j}}{(2j)!}.
\end{equation}
where
\begin{equation*}
  \kappa_j = \max_{t\in[0,\infty)} t^{j-t}.
\end{equation*}
In Appendix \ref{App:kappa} we show that
\begin{equation*}
  \frac{\kappa_{j-\varepsilon}}{\kappa_j} \; \underset{j\to\infty}{\text{\LARGE $\sim$}} \; \left(\frac{\ln j}{j}\right)^\varepsilon,
\end{equation*}
and, hence, the radius of convergence of $\sum_{j=1}^\infty b_jz^j$ is
\begin{eqnarray*}
\lim_{j\to\infty} \frac{b_{j-1}}{b_j} &=& \frac{1}{36} \lim_{j\to\infty} (2j)(2j-1)\frac{\kappa_{2j-2}}{\kappa_{2j}}\\
&=& \frac{1}{9} \lim_{j\to\infty} (\ln j)^2 = \infty.
\end{eqnarray*}
As we pointed out previously, this proves at the same time that $\rho(t)$ is defined for any value of $t$ and that $\rho(t)$ gives at any time $t$ the asymptotic of $\rho^{(L)}(t)$ for large $L$.

\subsubsection*{Convergence rate to the universal behavior. }
Our procedure has the advantage of making it possible to estimate the rate of convergence of $\rho^{(L)}(t)\xrightarrow{L\to\infty}\rho(t)$. From Eqs. \ref{Eqn:conv} and \ref{Eqn:b}, we obtain
\begin{equation*}
  |\rho(t)-\rho^{(L)}(t)| \leq \frac{2}{3} \sum_{j=L}^\infty \frac{\kappa_{2j}}{(2j)!} \, 6^{2j} |t|^{2j}.
\end{equation*}
Appendix \ref{App:kappa} proves that
\begin{equation*}
  \frac{\kappa_j}{j!} < \frac{1}{\omega_1\omega_2\cdots\omega_j},
\end{equation*}
where $\omega_j$ is the solution of $\omega_j+\ln\omega_j=1+\ln j$, which increases logarithmically with $j$ so that $\omega_j \sim \ln j$. Therefore,
\begin{equation*}
  |\rho(t)-\rho^{(L)}(t)| < \frac{2}{3} \sum_{j=L}^\infty \frac{6^{2j} |t|^{2j}}{\omega_1\omega_2\cdots\omega_{2j}},
\end{equation*}
which means that an upper bound for the error $|\rho(t)-\rho^{(L)}(t)|$ is given by
\begin{equation*}
  E^{(L)}(t) = \frac{2}{3} \frac{6^{2L} |t|^{2L}}{\omega_1\cdots\omega_{2L}} \left( 1 + \sum_{j=1}^\infty
\frac{6^{2j}|t|^{2j}}{\omega_{2L+1}\cdots\omega_{2L+2j}} \right).
\end{equation*}
Bearing in mind that $\omega_j$ grows with $j$, we find that the rate of consecutive error bounds satisfies
\begin{equation*}
  \frac{E^{(L)}(t)}{E^{(L-1)}(t)} < \frac{36|t|^2}{\omega_{2L-1}\omega_{2L}} \; \underset{L\to\infty}{\text{\LARGE $\sim$}} \;
\frac{36|t|^2}{\ln(2L-1)\ln(2L)},
\end{equation*}
i.e, the convergence $\rho^{(L)}(t) \xrightarrow{L\to\infty}\rho(t)$ is, at least, logarithmic.

The previous mathematical results suggest the following physical interpretation: For large $t$, the deviation from the universal behavior defined by $\rho(t)$ is related to the finiteness of the lattice. At the beginning of the evolution, the boundary conditions have little influence on the dynamics because the predominant interaction is the one between nearby atoms. Mathematically, the first coefficients of the power expansion of $\langle n_k(t) \rangle$ depend only on terms $H_l$ with $l$ close to $k$, so that the evolution is 'ignorant' of the boundaries. This is the origin of the universality observed for short times in Fig. \ref{Fig:universalregime}b. When $t$ increases, higher degree terms of $\langle n_k(t) \rangle$, which include contributions from $H_l$ with $l$ far from $k$, become significant. This means that the (indirect or mediated) interactions between distant atoms become more important as time passes. Finally, when $t$ is so large that the mediated interaction between atoms on sites at a distance of $L$ or more sites becomes relevant, the universality is lost. All this is a consequence of the fact that there is a finite speed at which information propagates, which is expected from the local interactions that govern the dynamics of the Rydberg gas \cite{Lieb72}.

\begin{figure}[h!!]
\includegraphics[width=\columnwidth]{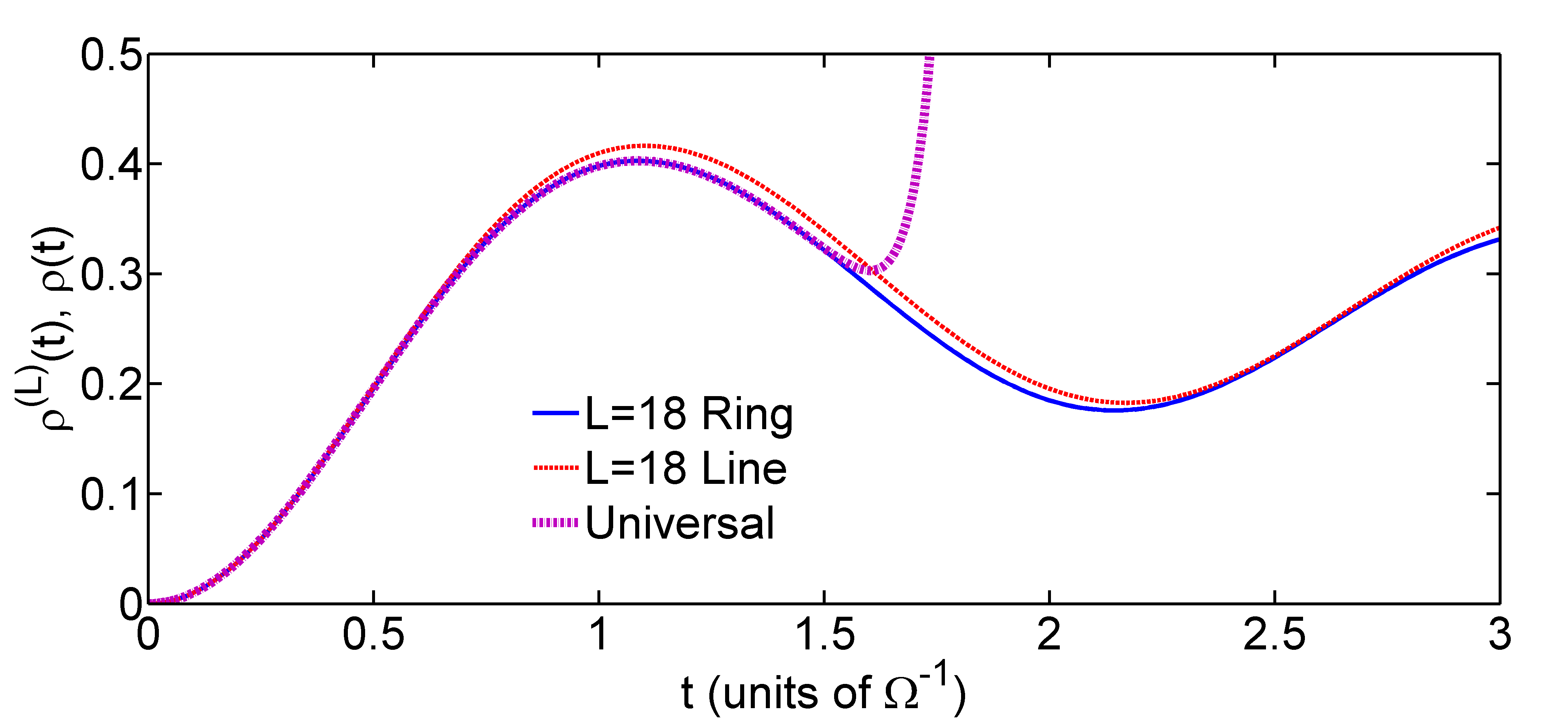}
\caption{(Color online) Comparison between the short-time evolution of the Rydberg density for a ring (blue solid line) and a linear (red dotted line) lattice with $L=18$, and the universal Rydberg density $\sum_{j=1}^{17}c_jt^{2j}$ (magenta thick-dotted line).}
\label{Fig:t34}
\end{figure}

This also means that the actual length of the universal time interval shown in Fig. \ref{Fig:universalregime}b depends on $L$: The lattice size fixes the number of coefficients $c_j^{(L)}$ which are universal and hence the range of values of $t$ for which $\rho^{(L)}(t)$ will not differ from the universal Rydberg density $\rho(t)$. Indeed, like the number of universal coefficients $c_j$, the range of the universal regime must increase with the size of $L$, and during it $\rho^{(L')}(t) \approx \rho^{(L)}(t)$ for $L' \gtrsim L$. At this point, the fact that the convergence $\rho^{(L)}(t)\xrightarrow{L\to\infty}\rho(t)$ takes place for any $t$ means that we can just increase $L$ in order to extend the length of the universal time interval. For finite values of $L$, these ideal power expansions can be calculated numerically in an alternative way using matrix representations of $H^{(L)}$ and $n^{(L)}$. Such computations can be simplified by reducing the operators to the maximally symmetric subspace under translations and reflections (see Ref. \cite{Olmos09-1}). As an example, we have performed this calculation for $L=18$ obtaining all the universal coefficients until $j=17$. A comparison between this result and the numerical time-evolution of a system of the same size with closed and open boundary conditions is presented in \autoref{Fig:t34}. Here the universal regime still lies in the transient phase and does not extend to the steady state regime. This universal time will increase with increasing number of sites on the lattice and, consequently, number of universal coefficients.

\subsubsection*{Beyond the nearest neighbors blockade. } \label{Sub:bnnb}
We will consider now a blockade radius $R_b$ larger than the lattice spacing $a$. Suppose that $R_b=\lambda_ba$, such that the blockade range is $\Lambda_k=\{k-\lambda_b,\dots,k-1,k+1,\dots,k+\lambda_b\}$ for any site $k$ of the ring. This yields an effective Hamiltonian $H^{(L)} = \sum_{k=1}^L H_k$ with
\begin{equation}\label{Eqn:hamil_lambda}
  H_k = m_{k-\lambda_b} \cdots m_{k-1} \Delta_k \, m_{k+1} \cdots m_{k+\lambda_b}.
\end{equation}

For the Rydberg density, we have  shown in Appendix \ref{App:univ} that $\mathrm{ad}_{H^{(L)}}^j(n_k)$ and $c_j^{(L)}$ are independent of $L$ for $j\leq(L-1)/\lambda_b$. Thus, there exists a universal set of expansion coefficients $c_j=\lim_{L\to\infty}c_j^{(L)}$ and a universal Rydberg density $\rho(t)=\sum_{j=1}^\infty c_jt^{2j}$ just like in the case $\lambda_b=1$.
For $\lambda_b=2$ and $\lambda_b=3$, some of the universal coefficients $c_{j}$ are provided in \autoref{tb:coefficients}. The bounds given in \autoref{Eqn:b} for the coefficients of the Rydberg density, become now (see Appendix \ref{App:words}),
\begin{equation*}
  |c_j|,|c_j^{(L)}| \leq 2(6\lambda_b)^{2j-1} \frac{\kappa_{2j+1/\lambda_b-1}}{(2j)!},
\end{equation*}
which proves that the Rydberg density $\rho^{(L)}(t)$ converges for any value of $t$ to the universal Rydberg density $\rho(t)$, no matter the size of the blockade radius.

Apart form the first term $t^2$, which is completely universal, the universal Rydberg density for a one-dimensional lattice only depends on the blockade radius.

\subsection{Expectation values of other operators} \label{SSub:other-op}
So far, we have only studied the Rydberg density. However, these results can actually be generalized to other local observables. Let us consider, for instance, the density-density correlation function
\begin{equation*}
  g_2(d,t)=\frac{\langle n_k(t) n_{k+d}(t) \rangle}{\langle n_k(t) \rangle \langle n_{k+d}(t) \rangle}
\end{equation*}
and, in particular, the expectation value
\begin{equation*}
  \rho_d^{(L)}(t) = \langle n_k(t) n_{k+d}(t) \rangle,
\end{equation*}
with $d \geq 2$, since $\langle n_kn_{k+1}\rangle=0$ due to the blockade condition.

As in the case of $\rho^{(L)}(t)$, $\rho_d^{(L)}(t)$ is an entire function even in $t$ and vanishing at $t=0$ with power expansion
\begin{equation} \label{Eqn:corr}
\rho_d^{(L)}(t) = \sum_{j=1}^\infty c_{d,j}^{(L)} t^{2j},
\end{equation}
where
\begin{equation*}
c_{d,j}^{(L)} = \frac{(-1)^j}{(2j)!} \langle\mathrm{ad}_{H^{(L)}}^{2j}(n_kn_{k+d})\rangle.
\end{equation*}
For very short times, atoms on separated sites evolve approximately independently. Thus, the density-density correlation function converges to 1 as $t\to 0$. From the universal behavior $\langle n_k(t) \rangle \sim t^2$, we conclude  that $\langle n_k(t) n_{k+d}(t) \rangle \sim t^4$ should be universal too, so that $c_{d,1}^{(L)}=0$ and $c_{d,2}^{(L)}=1$ which is the first  non-zero contribution in \autoref{Eqn:corr} being universal. Indeed, this can be shown following similar commutator computations to those done in \autoref{Sec:general} for the $t^2$ universal behavior of the Rydberg density.

Concerning the universality of the rest of the coefficients $c_{d,j}^{(L)}$, it is proven in Appendix \ref{App:univ} that, analogously to the case of the Rydberg density, $\mathrm{ad}_{H^{(L)}}^j(n_kn_{k+d})$ and $c_{d,j}^{(L)}$ are independent of $L$ for $j \leq L-d$. This gives rise to a set of universal coefficients $c_{d,j}$, some of which are given in \autoref{tb:coefficients} for $d=2$ and $3$, that allow us to construct the universal function.

Some of the previous findings can be extended to any observable. To formalize this statement we consider observables that are independent of the lattice size $L$ and that are constructed as a sum of \emph{words}. A word is defined as an operator $A = x_{k_1} x_{k_2} \cdots x_{k_j}$ with $k_1 < k_2 < \cdots < k_j$, built using the \emph{letters} $x_l\in\{r_l,r_l^\dag,m_l,n_l\}$. Each of these words is characterized by its length $\ell(A)=k_j-k_1+1$ and its  number of single letters $x_l\in\{r_l,r_l^\dag\}$, $s(A)$. Let us consider the expectation value of a word $A$ with length $\ell=\ell(A)$ and number of single letters $s=s(A)$,
\begin{equation*}
  \rho_A^{(L)}(t) = \langle A(t) \rangle.
\end{equation*}
Since the relation $(-1)^nA(-1)^n=(-1)^sA$ is accomplished, then $(-1)^nA(t)(-1)^n=(-1)^sA(-t)$ at any time. Hence, $\rho_A^{(L)}(-t) = (-1)^s \rho_A(t)$, i.e., the parity of $\rho_A^{(L)}(t)$ is determined by the number of single letters of $A$. As a consequence, the entire function $\rho_A^{(L)}(t)$ can be written as
\begin{equation*}
\rho_A^{(L)}(t) = \sum_{j=0}^\infty c_{A,j}^{(L)} t^j,
\qquad
c_{A,j}^{(L)} = \frac{i^j}{j!} \langle \mathrm{ad}_{H^{(L)}}^j(A) \rangle
\end{equation*}
where only powers with the same parity as $s$ contribute.

In this general situation, we can ensure that $\mathrm{ad}_{H^{(L)}}^j(A)$ and $c_{A,j}^{(L)}$ are universal at least for $j\leq (L-\ell)/2$ (see Appendix \ref{App:univ}), and we denote by $c_{A,j}$ these universal coefficients. Furthermore, the bounds
\begin{equation*}
|c_{A,j}|,|c_{A,j}^{(L)}| \leq 12^j \frac{\kappa_{j+\ell/2-1}}{j!},
\end{equation*}
obtained in Appendix \ref{App:words}, together with the asymptotics of $\kappa_j$ given in Appendix \ref{App:kappa} allow us to prove that
\begin{equation*}
\rho_A(t) = \sum_{j=0}^\infty c_{A,j}t^j
\end{equation*}
defines an entire function which picks up the universality of $\rho_A^{(L)}(t)$:
$\rho_A(t)\xrightarrow{L\to\infty}\rho_A^{(L)}(t)$ for any $t$.

This result shows that the universality observed in the time evolution of the Rydberg density is not caused by the specific form of the corresponding operator, but rather constitutes a general feature of this quantum system.

For a blockade radius larger than the lattice spacing, it has been shown in Appendix \ref{App:univ} that $\mathrm{ad}_{H^{(L)}}^j(A)$ and $c_{A,j}^{(L)}$ are universal at least for $j \leq (L-\ell)/(2\lambda_b)$, giving sense to the universal coefficients $c_{A,j}^{(L)}\xrightarrow{L\to\infty}c_{A,j}$ and the universal expectation value $\rho_A(t)=\sum_{j=1}^\infty c_{A,j}t^j$. Finally, the bounds
\begin{equation}\label{Eqn:cA}
  |c_{A,j}|,|c_{A,j}^{(L)}| \leq (12\lambda_b)^j \frac{\kappa_{j+\ell/(2\lambda_b)-1}}{j!}=: b_j,
\end{equation}
obtained in Appendix \ref{App:words} show that $\rho_A^{(L)}(t)\xrightarrow{L\to\infty}\rho_A(t)$ for any $t$.

\subsubsection*{Convergence rate to the universal behavior. }
We have seen that, in a ring under nearest neighbor blockade, the convergence of the expectation value of the Rydberg density to its universal counterpart is at least logarithmic. This estimate is generalized here to any other translation invariant or $L$-independent operator $A$, under any blockade range $\lambda_b$. Since $c_{A,j}^{(L)}=c_{A,j}$ for $j \leq (L-\ell)/(2\lambda_b)$, using the bounds \ref{Eqn:cA} in this general case, we find that
$$
|\rho_A(t)-\rho_A^{(L)}(t)| \leq 2 \! \sum_{j > \frac{L-\ell}{2\lambda_b}} b_j |t|^j =: E^{(L)}(t).
$$
In order to obtain the rate of consecutive error bounds $E^{(L)}$ and $E^{(L+2\lambda_b)}$ we need to calculate
$$
\frac{b_{j+1}}{b_j} = 12\lambda_b \frac{1}{j+1} \frac{\kappa_{j+\ell/(2\lambda_b)}}{\kappa_{j+\ell/(2\lambda_b)-1}}.
$$
From Appendix \ref{App:kappa} we get the following inequality
$$
\frac{b_{j+1}}{b_j} \leq 12\lambda_b \frac{j+\frac{\ell}{2\lambda_b}}{j+1} \frac{1}{\omega_{j+\frac{\ell}{2\lambda_b}}},
$$
with $\omega_k$ being the solution of $\omega_k+\ln\omega_k=1+\ln k$. Since $\omega_k$ is increasing in $k$, then
$$
\frac{b_{j+1}}{b_j} < 12\lambda_b \max\left\{1,\frac{L}{L-\ell+2\lambda_b}\right\} \frac{1}{\omega_{\frac{L}{2\lambda_b}}}, \quad
j>\frac{L-\ell}{2\lambda_b},
$$
which leads us to
$$
\begin{aligned}
E^{(L+2\lambda_b)}(t) & = 2 \! \sum_{j > \frac{L-\ell}{2\lambda_b}} b_{j+1} |t|^{j+1}
\\
& < 12\lambda_b \max\left\{1,\frac{L}{L-\ell+2\lambda_b}\right\} \frac{|t|}{\omega_{\frac{L}{2\lambda_b}}} E^{(L)}(t).
\end{aligned}
$$
Finally, bearing in mind that $\omega_k \sim \ln k$, we obtain the upper bound for the ratio
$$
\begin{aligned}
\frac{E^{(L+2\lambda_b)}(t)}{E^{(L)}(t)} & < 12\lambda_b \max\left\{1,\frac{L}{L-\ell+2\lambda_b}\right\} \frac{|t|}{\omega_{\frac{L}{2\lambda_b}}}
\\
& \underset{L\to\infty}{\text{\LARGE $\sim$}} \; 12\lambda_b \, \frac{|t|}{\ln\left(\frac{L}{2\lambda_b}\right)},
\end{aligned}
$$
which again shows that the convergence is at least logarithmic.

In summary, the universality of the evolution of the Rydberg density under nearest neighbors blockade also holds for the evolution of the expectation value of any other operator with an $L$-independent shape as well as for any operator that is invariant under translations, and for any blockade radius.
In addition, let us emphasize that the ring lattice is not merely a useful model to simplify the computations due to the symmetries, but also allows us to obtain the exact coefficients of the power expansion. That is, it grants us access to the universal behavior of the system independently of the lattice geometry, as it is discussed in the next Section.

\section{Linear lattice - boundary effects} \label{Sec:seg}
We will now investigate a one-dimensional lattice with open boundary conditions. Considering again nearest neighbors blockade, the effective Hamiltonian $H^{(L)}$ for a lattice with $L$ sites reads as in \autoref{Eqn:Hring} but with $m_0=m_{L+1}=1$. The analysis of the short time dynamics is similar to the ring lattice but the boundary effects are more pronounced. The reason is that, contrary to the ring, the atoms located at the edges of the line 'notice' the boundary conditions at any time. However, the theoretical approach followed previously in case of the ring can be extended also to this case.

We consider the expectation value of the Rydberg density  $\rho^{(L)}(t)=\langle n^{(L)}(t)\rangle/L$. It can be expanded as in expression \ref{Eqn:rho_L}, but the \autoref{Eqn:cj} for the coefficient $c_j^{(L)}$ of $t^{2j}$ is no longer valid because the linear lattice has no translation invariance. Only the invariance under reflection remains, so that the contributions of the sites $k$ and $L-k$ coincide. Therefore, in the coefficients
\begin{equation*}
  c_j^{(L)} = \frac{1}{L} \frac{(-1)^j}{(2j)!} \sum_{k=1}^L \langle \mathrm{ad}_{H^{(L)}}^{2j}(n_k) \rangle,
\end{equation*}
the contribution of each site $k$ must be computed independently.

The first coefficient $c_1^{(L)}$ is proportional to the expectation value $\langle \mathrm{ad}_{H^{(L)}}^2(n_k) \rangle$. First, we need $\mathrm{ad}_{H^{(L)}}^1(n_k) = [H_k,n_k]$, and to calculate it we follow the same route as in the ring when $k\neq1,L$, since for $k=1$ and $L$, one has $\mathrm{ad}_{H^{(L)}}^1(n_1)= \Delta_1 m_2$ and $\mathrm{ad}_{H^{(L)}}^1(n_L)= m_{L-1} \Delta_L$, respectively. Generalizing this for any $k$, it yields
\begin{equation*}
  \mathrm{ad}_{H^{(L)}}^1(n_k) = m_{k-1} \Delta_k m_{k+1}, \qquad m_0=m_{L+1}=1.
\end{equation*}
In the same spirit, one can write the operator $\mathrm{ad}_{H^{(L)}}^2(n_k) = [H^{(k-1)}+H^{(k)}+H^{(k+1)},\mathrm{ad}_{H_N}^1(n_k)]$ with $m_0=m_{L+1}=1$, and $m_{-1}=m_{L+2}=0$, so that \autoref{Eqn:ad2} remains valid for any $k$. From here we obtain again that $\langle \mathrm{ad}_{H^{(L)}}^2(n_k) \rangle =2\langle \delta_k \rangle = -2$ for any $k$ and
\begin{equation*}
  c_1^{(L)} = \frac{-1}{2L} \sum_{k=1}^L \langle \mathrm{ad}_{H^{(L)}}^2(n_k) \rangle = \frac{-1}{2L} (-2L) = 1.
\end{equation*}
This result was expected because we know that the $t^2$ behavior of the expectation value for the Rydberg density is independent of the geometry of the lattice.

The previous arguments show that the coefficients of the power series expansion of the Rydberg density on the linear lattice can be inferred from the $\mathrm{ad}_{H^{(L)}}^j(n_k)$ on the ring by setting $m_0=m_{L+1}=1$ and $m_k=0$ if $k<0$ or $k>L+1$. This rule not only simplifies the calculation but at the same time provides a direct comparison with the ring and its universal Rydberg density. Following this recipe we find that the next two coefficients are
\begin{equation*}
  c_2^{(L)}= -\left(1-\frac{2}{3L}\right)\qquad\mathrm{and}\qquad c_3^{(L)} =\frac{3}{5}\left(1-\frac{38}{27L}\right).
\end{equation*}
Hence, except for $c_1^{(L)}$, the coefficients $c_j^{(L)}$ are not universal, even for large $L$. However, for $L$ sufficiently large, all sites give the same contribution to the $j$-th coefficient, except for the $j-1$ sites counted from either end of the lattice. For each of these boundary sites, a fraction $p_1,\dots,p_{j-1}$ of the universal contribution is missing due to the condition $m_k=0$ if $k<0$ or $k>L+1$. Hence, we can rewrite
\begin{equation*}
  c_j^{(L)} = \frac{c_j}{L} \left(L - 2p_1 - \cdots - 2p_{j-1}\right) = c_j\left(1 - \frac{q_j}{L}\right),
\end{equation*}
where $q_j=2(p_1+\cdots+p_{j-1})$. Although the coefficients $c_j^{(L)}$ of the open lattice depend on its size $L$, the values of $q_j$ are $L$-independent provided that $L$ is large enough compared with $j$, in particular if $j\leq L+1$. This result can be checked with an alternative computation of $\mathrm{ad}_{H^{(L)}}^j(n^{(L)})$ for concrete values of $L$, using the matrix representation of $H^{(L)}$ and $n^{(L)}$ developed in Appendix \ref{App:matrixsegment}. The first universal values of $q_j$ obtained with these calculations are given in \autoref{tb:coefficients}

In general, there is no strict universality in the coefficients of the power expansion for $\rho^{(L)}(t)$ on the line. Nevertheless, as for a ring lattice, $c_j^{(L)}\xrightarrow{L\to\infty}c_j$ because $c_j^{(L)}=c_j+O(1/L)$. Furthermore, $\rho^{(L)}(t)$ also converges for any $t$ to the universal Rydberg density $\rho(t)$. The reason is that this convergence in the ring is simply a consequence of a bound for $c_j^{(L)}$ given by an overestimation of the number of \emph{words} in $\mathrm{ad}_{H^{(L)}}^{2j}(n_k)$ contributing to $c_j^{(L)}$ (see Appendix \ref{App:words}). Since the linear lattice reduces the number of such words due to the conditions $m_k=0$ for $k<0$ and $k>L+1$, the bound holds here too. As a consequence, we also expect the presence of a universal short time regime with a range that increases with the size of the system. A comparison between the Rydberg density for a $L=18$ linear lattice and the universal Rydberg density including terms with $j\le17$, i.e., $\sum_{j=1}^{17}c_jt^{2j}$, is presented in \autoref{Fig:t34}.

These results can be generalized to other operators and other blockade ranges.

\section{Multidimensional lattices} \label{Sec:dim}
For completeness, we provide some ideas on the generalization of these results to systems of higher dimensions. The analogue of a one-dimensional ring lattice of size $L$ in higher dimensions is a toroidal lattice defined by the periodic boundary conditions $k + Le_j \equiv k$ for any spatial basis vector $e_j$. The coefficients $c_j^{(L)}$ of the Rydberg density will depend not only on the blockade radius but also on the dimension of the system. Higher dimensional lattices yield different quantitative results for non-integer values of $\lambda_b$ due to the presence of neighbors at non-integer distances, absent in the one-dimensional case. Nevertheless, the short time behavior will be qualitatively similar: The first Taylor coefficients of the expectation value of a ($L$-independent or translationally invariant) operator are universal for $L$ large enough because the corresponding commutators involve sites with a relative distance not greater than $L$. Furthermore, the number of universal coefficients increases linearly with $L$. As in the one-dimensional case, these universality properties are affected by non-periodic boundary conditions (e.g., square and cubic lattices): The coefficients differ from the universal ones by $O(1/L)$ terms because, for large $L$, the number of boundary sites is of order $1/L$ compared with the total number of sites of the lattice.

Hence, we can at least formally define a power series expansion that represents the universal time evolution of expectation values of operators such as the Rydberg density. However, the reasoning used earlier in this paper to prove that this series actually defines a function (i.e., that the power series with the universal coefficients is convergent) does not apply to dimensions larger than one. The cornerstone of such methods is the bound obtained for the Taylor coefficients by overestimating the \emph{words} generated when applying $\mathrm{ad}_{H^{(L)}}^{j}$ to the corresponding operator (see Appendix \ref{App:words}). The number of such words grows exponentially with the dimension, which renders the techniques useless in dimension 2 or higher. This, of course, does not mean that there exists no such function as the universal Rydberg density in higher dimensions. In fact, its existence was proven previously in Ref. \cite{Nachtergaele06} although the method to calculate the corresponding universal function remains, to the best of our knowledge, unknown.

\section{Conclusions} \label{Sec:con}
In this work we have shown that expectation values of observables in a one-dimensional Rydberg lattice gas converge to  universal time dependent functions in the thermodynamic limit. This convergence takes place for any time, for any size-independent or translationally invariant observable, for any blockade radius, and for open and periodic boundary conditions. From a mathematical point of view, the universality can be understood as an asymptotic behavior when the size of the lattice goes to infinity. The expectation values of the considered operators converge (at least logarithmically) to some universal ones which could be associated to a model living in an infinite lattice. We have also shown how to obtain the expansion coefficients of these universal functions by investigating the dynamics on a ring lattice. The difficulty in the analysis of this asymptotics stems from the fact that such infinite lattice model is not well defined since the corresponding Hamiltonian has no meaning as an operator on a Hilbert space. Higher dimensional lattices also share in general these properties and the expectation values actually converge to universal ones as the lattice size increases. The universal expectation values given this asymptotics depend only on the operator at hand, the blockade radius, and the dimension of the lattice, but not on its particular geometry.

From the practical point of view the results presented in this paper mean that even the simulation of small systems allow to gain accurate insights into the behavior of macroscopic many-body systems, provided that the simulation time is sufficiently short, so that boundary effects are negligible. This has recently been exploited in the Ref. \cite{Gualdi11} which discusses the simulation of open quantum systems with local system-bath coupling.

\begin{acknowledgments}

L. Vel\'azquez wants to thank the Departament of Atomic, Molecular and Nuclear Physics from the University of Granada the hospitality during a stay where some parts of this work were done. The research of L. Vel\'azquez was partially supported by the research projects MTM2008-06689-C02-01 and MTM2011-28952-C02-01 from the Ministry of Science and Innovation of Spain and the European Regional Development Fund (ERDF), and by Project E-64 of Diputaci\'on General de Arag\'on (Spain). I. Lesanovsky acknowledges support by EPRSC. B. Olmos acknowledges funding by Fundaci\'on Ram\'on Areces. R. Gonz\'alez-F\'erez acknowledges financial support by the Spanish project FIS2011-24540 (MICINN) as well as by the Grants FQM-2445 and FQM-4643 (Junta de Andaluc\'{\i}a), she belongs to the Andalusian research group FQM-207.

\end{acknowledgments}

\appendix

\section{Universality on the ring} \label{App:univ}

\subsection{Rydberg density}

We will prove here that $\mathrm{ad}_{H^{(L)}}^j(n_k)$ and the coefficients $c_j^{(L)}$ of the Rydberg density are universal for $j \leq L-1$. The proof of these results relies on the following basic facts:

\medskip

\noindent {\bf (P1)} For $1 \leq j \leq L-1$, $\mathrm{ad}_{H^{(L)}}^j(n_k)$ is a sum of \emph{words} $x=x_p x_{p+1} \cdots x_{p+N}$ which are products of \emph{letters} $x_l\in\{r_l,r_l^\dag,m_l,n_l\}$. The initial and final letters are always $x_p=m_p$ and $x_{p+N}=m_{p+N}$. The length $\ell(x)=N+1$ of each word $x$ is not greater than $j+2$ and $p<k<p+N$.

\medskip

\noindent {\bf (P2)} We  denote by  $s(x)$  the number of \emph{single} letters $x_l\in\{r_l,r_l^\dag\}$ in the word $x=x_p x_{p+1} \cdots x_{p+N}$. If $2 \leq \ell(x) \leq L$ and $x_p=m_p$, $x_{p+N}=m_{p+N}$, the commutator $[H^{(L)},x]$ is a sum of words $x'$ with $\ell(x') \geq \ell(x)$ and $s(x')=s(x)\pm1$. Furthermore,
$$
\ell(x')>\ell(x) \; \Rightarrow \; \ell(x')=\ell(x)+1, \; s(x')=s(x)+1.
$$

\begin{proof}
We will prove (P1) and (P2) by induction on $j$. Equations \ref{Eqn:ad1} and  \ref{Eqn:ad2} show that they hold for $j=1$ and $L\geq2$, and $j=2$ with $L\geq3$, respectively. For $j=1$, $\mathrm{ad}_{H^{(L)}}^1(n_k)$ is composed by two words, with $\ell(x)=3$ and $s(x)=1$. For $j=2$, we find in $\mathrm{ad}_{H^{(L)}}^2(n_k)$ words with $\ell(x)=3$ and $s(x)=0$, and others with $\ell(x)=4$ and $s(x)=2$. Assuming that they are true for an index $j-1 \leq L-2$, we will prove that they also hold for $j$.
Under the induction hypothesis, $\mathrm{ad}_{H^{(L)}}^{j}(n_k)$ should be a sum of the words obtained from $[H^{(L)},x]$, where $x=m_px_{p+1} \cdots x_{p+N-1}m_{p+N}$ are words of $\mathrm{ad}_{H^{(L)}}^{j-1}(n_k)$, which must satisfy $p < k < {p+N}$ and $\ell(x)\leq j+1$.

Since $\ell(x) \leq j+1 \leq L$, the word $x$ contains at most as many letters as terms $H_l$ are in $H^{(L)}$. Hence, $[H^{(L)},x]=[\sum_{l=p}^{p+N}H_l,x]$ where the sum runs over different indices mod($L$). This fact, together with \autoref{Eqn:mr}, yields the decomposition
\begin{equation} \label{Eqn:Hx}
\begin{aligned}
\kern-5pt \textstyle \sum_{l=p}^{p+N}H_l x & = m_{p-1} \Delta_pm_p m_{p+1}x_{p+1} x_{p+2} \cdots x_{p+N-1} m_{p+N}
\\
& \textstyle + m_p \left( \sum_{l=p+1}^{p+N-1}H_l \right) x_{p+1} \cdots x_{p+N-1} m_{p+N}
\\
& + m_p x_{p+1} \cdots x_{p+N-2} m_{p+N-1}x_{p+N-1} \\
&\text{ } \Delta_{p+N}m_{p+N} m_{p+N+1}
\\
& = m_{p-1} r_p^\dag \tilde{x}_{p+1} x_{p+2} \cdots x_{p+N-1} m_{p+N}
\\
& \textstyle + m_p \left(\sum_y y_{p+1} \cdots y_{p+N-1}\right) m_{p+N}
\\
& + m_p x_{p+1} \cdots x_{p+N-2} \tilde{x}_{p+N-1} r_{p+N}^\dag m_{p+N+1}.
\end{aligned}
\end{equation}
where $ \tilde{x}_{l}=m_lx_l$ are new letters and $\sum_y$ is a sum over the words $y=y_{p+1} \cdots y_{p+N-1}$. This equation shows that $\sum_{l=p}^{p+N}H_lx$ is a sum of words $x'$ with operators $m_l$
at the right and left extremes. The words have length $\ell(x')=\ell(x)$ or $\ell(x')=\ell(x)+1$, and thus $\ell(x')\leq j+2$. Since we have chosen $j-1 \leq L-2$, then $j\leq L-1$. For $\sum_{l=p}^{p+N} xH_l$, an analogous  relation to \ref{Eqn:Hx} satisfying the same  properties is obtained. Therefore, the statement (P1) is proven.

In addition, the multiplication by the factors $m_{l}$ does not alter the number of single letters but the words $x_l$ and $\Delta_lx_l$ always differ by one in the number of single letters (see \autoref{Eqn:mr}). Hence, $\sum_{l=p}^{p+N}H_lx$ is a sum of words $x'$ with $s(x')=s(x)\pm1$. Furthermore, if $\ell(x')=\ell(x)+1$, then $s(x')=s(x)+1$. A similar analysis for $\sum_{l=p}^{p+N}xH_l$ gives rise to the same results, and therefore (P2)
is proven.
\end{proof}

\subsubsection{Universality of $\mathrm{ad}_{H^{(L)}}^j(n_k)$ for $j \leq L-1$. } \label{SSub:uni-ad}

Result (P1) ensures that, for $j \leq L$, $\mathrm{ad}_{H^{(L)}}^j(n_k)=[H^{(L)},\mathrm{ad}_{H^{(L)}}^{j-1}(n_k)]$ involves only commutators of $H^{(L)}$ with words $x = m_p x_{p+1}\cdots x_{p+N-1} m_{p+N}$ of length not greater than $j+1$. Therefore, for $j\leq L-1$, $\mathrm{ad}_{H^{(L)}}^j(n_k)$ is a sum of terms $[\sum_{l=p}^{p+N}H_l,x]$ with different indices $l$ mod($L$), that makes the calculation universal since it is determined only by the universal algebra of $r_l$ and $r_l^\dag$.

For $j \geq L$, if we compute the commutators without simplifying the factors at the extremes with the same index mod($L$), i.e., as in \autoref{Eqn:ad2L=2}, $\mathrm{ad}_{H^{(L)}}^j(n_k)$ becomes a sum of words already present in the universal case, but with some of them missing. This is due to the fact that the words $x = x_p \cdots x_{p+N}$ from such a computation can have letters with indices which are equal mod($L$). Then, when doing $[H_l,x]$, in $H_lx$, $H_l$  acts from the left on the first factors $x_{l_1-1}x_{l_1}x_{l_1+1}$ such that $l_1=l$ mod($L$), whereas in $xH_l$, $H_l$ acts from the right on the first factors $x_{l_2-1}x_{l_2}x_{l_2+1}$ such that $l_2=l$ mod($L$). That is, $[H_l,x]$ reproduces the terms $H_{l_1}x$ and $-xH_{l_2}$ of the universal
computation of $[H_{l_1},x]$ and $[H_{l_2},x]$, but the terms $-xH_{l_1}$ and $H_{l_2}x$ are missing, together with the full commutators $[H_{l'},x]$ for any index $l'=l$ mod($L$) such that $l_1 < l' < l_2$.

\subsubsection{Universality of $c_j^{(L)}$ for $j \leq L-1$. } \label{SSub:uni-c}

As a consequence of (P2), $\mathrm{ad}_{H^{(L)}}^j(n_k)$ is a sum of words whose number of single letters has the same parity as $j$. The only words of $\mathrm{ad}_{H^{(L)}}^j(n_k)$ that give a non-zero contribution to $\langle\mathrm{ad}_{H^{(L)}}^j(n_k)\rangle$ are products of $m_l$-operators. Thus, for those words $s(x)=0$ and, hence, they appear only when $j$ is even.

Any word $x^{(2j)}$ of $\mathrm{ad}_{H^{(L)}}^{2j}(n_k)$ comes from a previous one $x^{(2j-1)}$ of $\mathrm{ad}_{H^{(L)}}^{2j-1}(n_k)$ when doing $[H^{(L)},x^{(2j-1)}]$ and, analogously, $x^{(2j-1)}$ comes from $x^{(2j-2)}$ of $\mathrm{ad}_{H^{(L)}}^{2j-2}(n_k)$. Following this procedure, each word $x^{(2j)}$ of $\mathrm{ad}_{H^{(L)}}^{2j}(n_k)$ has a \emph{history},
$$
n_k=x^{(0)} \to x^{(1)} \to \cdots \to x^{(2j-1)} \to x^{(2j)}=x,
$$
where $x^{(g)}$, at the \emph{generation} $g$, is the word of $\mathrm{ad}_{H^{(L)}}^g(n_k)$ giving rise to $x^{(g+1)}$ (among other words) through $[H^{(L)},x^{(g)}]$.

Result (P2) states that $\ell(x^{(g+1)})\geq\ell(x^{(g)})$ and $s(x^{(g+1)})=s(x^{(g)})\pm1$. In addition, (P2) ensures that each step $x^{(g)} \to x^{(g+1)}$ with $g \geq 1$ can increase the length of the word at most by one unit, and, if this happens, the number of single letters must increase by one unit. In the first step $x^{(0)} \to x^{(1)}$, the length is increased by two units and the number of single letters by one unit.

Since $x^{(0)}=n_k$, the history of a word $x=x^{(2j)}$ which is a product of  $m_l$ operators must have the same number of steps rising and lowering the number of single letters, given that $s(x)=0$. Hence, at most $j$ steps of the history of $x$ can increase the length and, as a consequence, the intermediate words $x^{(g)}$ must accomplish $\ell(x^{(g)})\leq j+2$. Let us show by induction on $g$ that the history of such a word $x$ must be universal for $j\leq L-1$: We assume that $x^{(g)}$ is a universal word of $\mathrm{ad}_{H^{(L)}}^g(n_k)$. Since $\ell(x^{(g)}) \leq j+1$, analogously to the discussion of the universality of $\mathrm{ad}_{H^{(L)}}^j(n_k)$, we conclude that $[H^{(L)},x^{(g)}]$ is universal for $j \leq L-1$. 
As a consequence, the words $x$ of $\mathrm{ad}_{H^{(L)}}^{2j}(n_k)$ contributing to $\langle\mathrm{ad}_{H^{(L)}}^{2j}(n_k)\rangle$ are universal for $j \leq L-1$, and thus $c_j^{(L)}$ becomes universal for $j \leq L-1$.

\subsection{Other operators}

The analysis of $\langle n_k(t)n_{k+d}(t) \rangle$ requires to consider words $x = x_{l_1} x_{l_2} \cdots x_{l_i}$, $x_l \in \{r_l,r_l^\dag,m_l,n_l\}$, with indices $l_1 < l_2 < \cdots < l_i$ that could be non consecutive. We define their length $\ell(x)$ by $\ell(x)=l_i-l_1+1$.

At each step $\mathrm{ad}_{H^{(L)}}^{j-1}(n_kn_{k+d})\to\mathrm{ad}_{H^{(L)}}^j(n_kn_{k+d})$, the length of the words increases by one unit or remains invariant, and the number of single letters increases or decreases by one unit. In particular, the first step always increases the length, and, at any step, an enhancement of the length also implies an increase of the number of single letters. Hence, we obtain the following results for $j \leq L-d$: i) the operator $\mathrm{ad}_{H^{(L)}}^j(n_kn_{k+d})$ is a sum of words with initial and final letters of type $m_l$ or $n_l$, and with length between $d+1$ and $j+d+1$; ii) if $j\geq1$, the minimum length is $d+2$. Hence, the same reasoning followed for the Rydberg density proves that $\mathrm{ad}_{H^{(L)}}^j(n_kn_{k+d})$ and $c_{d,j}^{(L)}$ are independent of $L$ for $j \leq L-d$.

We analyze  now the general case of an operator $A$ which is a word of length $\ell$. All that we can ensure is that in each step $\mathrm{ad}_{H^{(L)}}^{j-1}(A) \to \mathrm{ad}_{H^{(L)}}^j(A)$ preserves the length of the words or increases it by one or two units, while the number of single letters increases or decreases by one unit. From here and following similar arguments as in  the previous cases, we find that $\mathrm{ad}_{H^{(L)}}^j(A)$ and $c_{A,j}^{(L)}$ are universal at least for $j\leq (L-\ell)/2$.

\subsection{Beyond nearest neighbors blockade}

Now, we consider a blockade radius $R_b = \lambda_b a$ with the Hamiltonian given in \autoref{Eqn:hamil_lambda}. Concerning the Rydberg density, if $j \leq (L-1)/\lambda_b$ and $j\geq1$, the operator $\mathrm{ad}_{H^{(L)}}^j(n_k)$ is a sum of words with left and right ends of type $m_{l+1} m_{l+2} \cdots m_{l+\lambda_b}$ and with length between $2\lambda_b+1$ and $\lambda_b(j+1)+1$. This is due to the following facts: i) each step $\mathrm{ad}_{H^{(L)}}^{j-1}(n_k) \to \mathrm{ad}_{H^{(L)}}^j(n_k)$ increases the length of the words by at most $\lambda_b$ units, except for the first one which increases it by $2\lambda_b$ units; ii) at any step, the number of single letters increases or decreases by one unit, and an increase of the length also implies an increase of the number of single letters. From this, similarly to the case of nearest neighbors blockade ($\lambda_b=1$), we find that $\mathrm{ad}_{H^{(L)}}^j(n_k)$ and $c_j^{(L)}$ are independent of $L$ for $j \leq (L-1)/\lambda_b$.

For an arbitrary operator $A$ of length $\ell$, each step $\mathrm{ad}_{H^{(L)}}^{j-1}(A) \to \mathrm{ad}_{H^{(L)}}^j(A)$ increases its length by at most $2\lambda_b$ units, while increasing or decreasing the number of single letters by one. This implies that $\mathrm{ad}_{H^{(L)}}^j(A)$ and $c_{A,j}^{(L)}$ are universal for $j \leq (L-\ell)/(2\lambda_b)$.

\section{Bounds for $c_{j}^{(L)}$ and $c_{A,j}^{(L)}$} \label{App:words}

\subsection{The case of the Rydberg density}

A very rough bound for $|c_j^{(L)}|$ is given by $1/(2j)!$ times the total number of words of $\mathrm{ad}_{H^{(L)}}^{2j}(n_k)$ because each one of its terms contributes to $\langle\mathrm{ad}_{H^{(L)}}^{2j}(n_k)\rangle$ with the value 1, -1 or 0. Since the number of words of $\mathrm{ad}_{H^{(L)}}^j(n_k)$ is non decreasing with $L$ and constant for $j \leq L-1$, we conclude that
$$
|c_j|, |c_j^{(L)}| \leq \frac{w_{2j}}{(2j)!},
$$
with $w_j$ being the number of words in $\mathrm{ad}_{H^{(L)}}^j(n_k)$ for $j\leq L-1$.

In order to count the number of words in $\mathrm{ad}_{H^{(L)}}^j(n_k)$, we start from $\mathrm{ad}_{H^{(L)}}^1(n_k)$, that has only 2 words of length 3, and make $j-1$ steps to arrive at $\mathrm{ad}_{H^{(L)}}^j(n_k)$. According to
\autoref{Eqn:Hx}, for any word $x$ of $\mathrm{ad}_{H^{(L)}}^g(n_k)$, the commutator $[H^{(L)},x]$ gives at most 4 words with length $\ell(x)+1$ and $2(\ell(x)-2)$ with length $\ell(x)$. Thus, if we consider only the histories
with exactly $l$ steps increasing the length of the words, there are $\binom{j-1}{l}$ possibilities for choosing them. Given a certain choice of these $l$ steps, the number of histories is bounded by $2\cdot4^l(2(l+1))^{j-1-l}$: Each of the $l$ steps increasing the length gives at most 4 new words for each old one; the remaining $j-1-l$ steps must preserve the length and any of them gives at most $2(\ell-2)$ new words for each old one of length $\ell$. This result is obtained taking into account that $l$ steps increasing the length give a word of length $l+3$, so at any intermediate step $\ell \leq l+3$.

Therefore, defining $\kappa_j = \max_{t\in[0,\infty)} t^{j-t}$, one obtains
\begin{equation} \label{Eqn:wj}
\begin{aligned}
w_j & \leq \sum_{l=0}^{j-1} \binom{j-1}{l} \, 2 \cdot 4^l (2(l+1))^{j-l-1}
\\
& \leq 2 \kappa_j \sum_{l=0}^{j-1} \binom{j-1}{l} 4^l 2^{j-1-l} = 2 \cdot 6^{j-1} \kappa_j.
\end{aligned}
\end{equation}
Using this result, we finally find that
$$
|c_j|,|c_j^{(L)}| \leq 2 \cdot 6^{2j-1} \frac{\kappa_{2j}}{(2j)!},
$$
which is the bound for the coefficients we were looking for.

\subsection{The case of a general word $A$}

Let us assume that $A$ is a word of length $\ell$. We split the $j$ steps of $A \to \mathrm{ad}_{H^{(L)}}^j(A)$ into histories with $l_1$ and $l_2$ steps increasing the length in one and two units, respectively. Then, the number of words $w_{A,j}$ of $\mathrm{ad}_{H^{(L)}}^j(A)$ can be bounded by
$$
\begin{aligned}
& w_{A,j} \leq \sum_{l_1+l_2 \leq j} \binom{j}{l_1 \; l_2} 4^{l_1+l_2} (2(l_1+2l_2+\ell-2))^{j-l_1-l_2}
\\
& \leq \kern-3pt \sum_{l_1+l_2 \leq j} \binom{j}{l_1 \; l_2} 4^j (l_1+l_2+\textstyle\frac{\ell}{2}-1)^{j-l_1-l_2} \leq  12^j
\kappa_{j+\frac{\ell}{2}-1}.
\end{aligned}
$$
The multinomial coefficient $\binom{j}{l_1 \; l_2}$ counts the number of ways of choosing $l_k$ steps increasing the length in $k$ units among the $j$ steps in $A \to \mathrm{ad}_{H^{(L)}}^j(A)$. Starting at $A$ with length $\ell$, each of the $l_k$ steps generates at most 4 new words from each old one. The remaining $j-l_1-l_2$ steps keep invariant the length and, thus, none of them yields more than $2(\ell'-2)$ new words for each old one of length $\ell'$. The bound follows from the fact that $\ell' \leq \ell+l_1+2l_2$.

Since $\langle\mathrm{ad}_{H^{(L)}}^j(A)\rangle$ is bounded by the total number of words of $\mathrm{ad}_{H^{(L)}}^j(A)$, one finally obtains
$$
|c_{A,j}|,|c_{A,j}^{(L)}| \leq \frac{w_{A,j}}{j!} \leq 12^j \frac{\kappa_{j+\ell/2-1}}{j!},
$$
as the bounds for the coefficients.

\subsection{Non nearest neighbors blockade}

For a blockade radius $R_b = \lambda_b a$, the bound for the number  of words  $w_j$ of $\mathrm{ad}_{H^{(L)}}^j(n_k)$ that generalizes \autoref{Eqn:wj} yields
$$
\begin{aligned}
w_j & \leq \sum_{l=0}^{j-1} \binom{j-1}{l} 2 (4\lambda_b)^l (2(\lambda_bl+1))^{j-l-1}
\\
& \leq 2 (6\lambda_b)^{j-1} \kappa_{j+\frac{1}{\lambda_b}-1}.
\end{aligned}
$$
Among the $j-1$ steps in $\mathrm{ad}_{H^{(L)}}^1(n_k) \to\mathrm{ad}_{H^{(L)}}^j(n_k)$ there are $\binom{j-1}{l}$ ways of choosing $l$ steps with increasing length. There are 2 words of length $2\lambda_b+1$ in $\mathrm{ad}_{H^{(L)}}^1(n_k)$. Each of the $l$ steps increasing the length gives at most $4\lambda_b$ new words from each old one. The remaining $j-1-l$ steps do not increase the length and, hence, none of them yields more than $2(\ell-2\lambda_b)$ new words for each old one of length $\ell$. This leads to the above inequality because $\ell \leq \lambda_bl+(2\lambda_b+1)$ when only $l$ steps increase the length. Then, the coefficients of the Rydberg density are bounded by
$$
|c_j|,|c_j^{(L)}| \leq 2(6\lambda_b)^{2j-1} \frac{\kappa_{2j+1/\lambda_b-1}}{(2j)!}.
$$

For a word $A$ of length $\ell$, a bound for the number of words $w_{A,j}$ of $\mathrm{ad}_{H^{(L)}}^j(A)$ follows by splitting the $j$ steps of $A \to \mathrm{ad}_{H^{(L)}}^j(A)$ into histories combining $l_k$ steps increasing the length of the words in $k$ units with $k=1,2,\dots,2\lambda_b$. This yields the following bound for $w_{A,j}$,
$$
\sum_{\text{\tiny$\displaystyle \sum_k$} \, l_k \leq j} \binom{j}{l_1 \; l_2 \; \cdots \; l_{2\lambda_b}} 4^{\text{\tiny$\displaystyle \sum_k$} \,
l_k} \big(2\big(\textstyle \sum_k kl_k + \ell - 2\lambda_b\big)\big)^{j - \text{\tiny$\displaystyle \sum_k$} \, l_k},
$$
and using the inequality $\sum_k kl_k + \ell - 2\lambda_b \leq 2\lambda_b \big(\sum_k l_k + \ell/2\lambda_b-1\big)$, we arrive at
$$
w_{A,j} \leq (12\lambda_b)^j \kappa_{j+\frac{\ell}{2\lambda_b}-1},
$$
and hence,
$$
|c_{A,j}|,|c_{A,j}^{(L)}| \leq (12\lambda_b)^j \frac{\kappa_{j+\ell/2\lambda_b-1}}{j!}.
$$

\section{Asymptotics of $\kappa_j$} \label{App:kappa}

Let $a$ be any positive real number and
$$
\kappa(a) = \max_{t\in[0,\infty)} f_a(t), \qquad f_a(t)=t^{a-t}.
$$
We will prove here that, for any $\varepsilon$,
\begin{equation*} 
\frac{\kappa(a-\varepsilon)}{\kappa(a)} \; \underset{a\to\infty}{\text{\LARGE $\sim$}} \; \left(\frac{\ln a}{a}\right)^\varepsilon.
\end{equation*}

First, $f_a(t)$ is a positive and $C^{\infty}$ function in $(0,\infty)$ such that $\lim_{t\to0}f_a(t)=\lim_{t\to\infty}f_a(t)=0$. This implies that $f_a(t)$ reaches its maximum at a point $\tau\in(0,\infty)$ such that $f'_a(t=\tau) =
f_a(\tau)(a/\tau-1-\ln \tau)=0$, i.e.,
\begin{equation} \label{Eqn:tau}
\frac{a}{\tau}-1-\ln \tau=0.
\end{equation}
The solution of this equation is unique for any $a$ because $g_a(t)=a/t-1-\ln t$ is decreasing in $(0,\infty)$. Then,
$$
\begin{gathered}
\kappa(a) = f_a(\tau) = \tau^{a-\tau} = \tau^{\tau\ln\tau} = e^{\sigma(a)},
\end{gathered}
$$
with $\sigma(a) = \tau(\ln\tau)^2$ and
\begin{equation} \label{Eqn:lim}
\frac{\kappa(a-\varepsilon)}{\kappa(a)}=\frac{1}{e^{\sigma(a)-\sigma(a-\varepsilon)}}.
\end{equation}

Both $\tau(a)$ and $\sigma(a)$ are $C^{\infty}$ functions of
$a\in(0,\infty)$ because $g_a(t)$ is $C^{\infty}$ for
$a,t\in(0,\infty)$ and $g'_a(t)<0$
in this interval. In addition,
$$
\begin{aligned}
& \tau' (a)= \frac{\tau}{a+\tau} = \left(\frac{a}{\tau}+1\right)^{-1}
= \frac{1}{2+\ln\tau},
\\
& \sigma' (a)= \tau'\ln\tau(2+\ln\tau) = \ln\tau,
\end{aligned}
$$
are positive functions for $a>0$ and $1$, respectively, hence
$\tau(a)$ and $\sigma(a)$ are increasing in the corresponding intervals:
$\tau'(a)=\tau/(a+\tau)>0$ for $a>0$ because $\tau>0$ in that interval
by construction; as a consequence, if $a>1$ then $\tau(a)>\tau(1)=1$ and
$\sigma'(a)>\ln\tau(1)=0$.

The mean value theorem yields
$$
\sigma(a)-\sigma(a-\varepsilon) = \varepsilon \sigma'(\xi_a) = \varepsilon \ln\tau(\xi_a), \quad \xi_a\in(a-\varepsilon,a),
$$
so that, in \autoref{Eqn:lim},
\begin{equation} \label{Eqn:kk}
\frac{\kappa(a-\varepsilon)}{\kappa(a)} = \frac{1}{(\tau(\xi_a))^\varepsilon}.
\end{equation}
Using \autoref{Eqn:tau}, we find that $\omega(a)=a/\tau(a)$ satisfies
\begin{equation*}
\omega + \ln\omega = 1 + \ln a,
\end{equation*}
which proves that $\omega(a)$ is increasing, $\omega(a)\xrightarrow{a\to\infty}\infty$ and that in that limit $\omega(a) \sim \omega(a)+ \ln \omega(a) \sim \ln a$. Therefore, \autoref{Eqn:kk} yields
$$
\frac{\kappa(a-\varepsilon)}{\kappa(a)} = \left(\frac{\omega(\xi_a)}{\xi_a}\right)^\varepsilon \sim \left(\frac{\ln a}{a}\right)^\varepsilon,
$$
where we have used that $\lim_{a\to\infty}(a/\xi_a)=1$.

In addition, since $\tau(a)$ is increasing and $\xi_a < a$, one obtains that
\begin{equation*}
\frac{\kappa(a-\varepsilon)}{\kappa(a)} > \frac{1}{(\tau(a))^\varepsilon} = \left(\frac{\omega(a)}{a}\right)^\varepsilon,
\end{equation*}
so that, finally, given that $\tau(1)=\kappa(1)=1$ and denoting $\kappa_n=\kappa(n)$ and $\omega_n=\omega(n)$ with $n=1,2,3,\dots$,
$$
\kappa_n = \frac{\kappa_n}{\kappa_{n-1}} \frac{\kappa_{n-1}}{\kappa_{n-2}} \cdots \frac{\kappa_2}{\kappa_1} <
\frac{n!}{\omega_1\omega_2\cdots\omega_n}.
$$

\section{Matrix representations on the linear lattice} \label{App:matrixsegment}

Consider a linear lattice of $L$ sites under nearest neighbors blockade. The Hilbert state space has dimension $2^L$ and a basis is given by the eigenvectors with Rydberg excitation numbers $n_1,\dots,n_L$, which we label with the corresponding eigenvalues. The absence of translational symmetry leaves only the reflection symmetry to reduce the degrees of freedom, but it is unable to eliminate even half of them. Instead of this, we will reduce the degrees of freedom working in the blockade subspace, spanned by those states without consecutive Rydberg excitations. A basis $B^{(L)}$ of the blockade subspace can be inductively constructed on $L$ as
\begin{equation} \label{Eqn:BL}
\begin{aligned}
& B^{(L)} = B^{(L-1)}\otimes|0\rangle \cup B^{(L-2)}\otimes|01\rangle,
\\
& B^{(1)}=\{|0\rangle,|1\rangle\}, \quad B^{(2)}=\{|00\rangle,|10\rangle,|01\rangle\},
\end{aligned}
\end{equation}
where $0$ and $1$ represent an atom in the ground and Rydberg state, respectively. The dimension $d^{(L)}$ of the blockade subspace satisfies
$$
d^{(L)} = d^{(L-1)} + d^{(L-2)}, \qquad d^{(1)}=2, \quad d^{(2)}=3,
$$
which defines the Fibonacci sequence, given explicitly by
$$
d^{(L)}=\frac{r^{L+2}-(-1/r)^{L+2}}{\sqrt{5}}, \qquad r=\frac{1+\sqrt{5}}{2}.
$$
Hence, the restriction to the blockade subspace reduces the degrees of freedom from $2^L$ to $d^{(L)} \sim \frac{r^{L+2}}{\sqrt{5}}$.

In the basis $B^{(L)}$, the matrix representation $\mathcal{N}^{(L)}$ of the total number of Rydberg excitations $n^{(L)}=\sum_{k=1}^{L}n_k$ is given by the recursion
$$
\begin{aligned}
& \mathcal{N}^{(L)} = \begin{pmatrix} \mathcal{N}^{(L-1)} & 0 \\ 0 & \mathcal{N}^{(L-2)}+I_{L-2} \end{pmatrix},
\\
& \mathcal{N}^{(1)} = \begin{pmatrix} 0 & 0 \\ 0 & 1 \end{pmatrix},
\quad
\mathcal{N}^{(2)} = \begin{pmatrix} 0 & 0 & 0 \\ 0 & 1 & 0 \\ 0 & 0 & 1 \end{pmatrix},
\end{aligned}
$$
where $I_i$ stands for the identity matrix of order $d^{(i)}$.

Concerning the Hamiltonian $H^{(L)}$, its matrix representation $\mathcal{H}^{(L)}$ in the basis $B^{(L)}$ is defined as follows:
$H^{(L)}$ transforms a state of $B^{(L)}$ into a sum of those states obtained changing a $1$ by $0$, or a $0$ by $1$ whenever there is no $1$ in the contiguous sites. Bearing in mind \autoref{Eqn:BL}, the action of $H^{(L)}$ on $B^{(L)}$ is determined by its action on $B^{(L-1)} \otimes |0\rangle$ and $B^{(L-2)} \otimes |01\rangle$. Writing $B^{(L-1)} \otimes |0\rangle = B^{(L-2)}\otimes|00\rangle \cup B^{(L-3)}\otimes|010\rangle$ shows that
$$
\begin{aligned}
H^{(L)}(B^{(L-1)} \otimes |0\rangle) = & \; H^{(L-1)}B^{(L-1)} \otimes |0\rangle
\\
& + (B^{(L-2)} \otimes |01\rangle,\underbrace{0,0,\dots,0}_{d^{(L-3)}}).
\end{aligned}
$$
On the other hand,
$$
H^{(L)} (B^{(L-2)} \otimes |01\rangle) = H^{(L-2)} B^{(L-2)} \otimes |01\rangle + B^{(L-2)} \otimes |00\rangle,
$$
where we remind that $B^{(L-2)} \otimes |00\rangle$ are the first $d^{(L-2)}$ states of $B^{(L)}$.

The previous identities translate into a recurrence for $\mathcal{H}^{(L)}$ given by
$$
\mathcal{H}^{(L)} = \begin{pmatrix} \mathcal{H}^{(L-1)} & \mathcal{K}^{(N)} \\ (\mathcal{K}^{(N)})^t & \mathcal{H}^{(L-2)} \end{pmatrix}, \quad
\mathcal{K}^{(N)} = \begin{pmatrix} I_{L-2} \\ 0_{L-3,L-2} \end{pmatrix},
$$
with $0_{i,j}$ being the null $d^{(i)} \times d^{(j)}$ matrix. This generates the matrix $\mathcal{H}^{(L)}$ for any $L$ starting from
$$
\mathcal{H}^{(1)} = \begin{pmatrix} 0 & 1 \\ 1 & 0 \end{pmatrix},
\qquad
\mathcal{H}^{(2)} = \begin{pmatrix} 0 & 1 & 1 \\ 1 & 0 & 0 \\ 1 & 0 & 0 \end{pmatrix}.
$$

Since $|\mathbf{0}\rangle$ is the first element of the basis $B^{(L)}$, the expectation value $\rho^{(L)}(t)=\frac{1}{L}\langle\mathbf{0}|e^{itH^{(L)}}n^{(L)}\,e^{itH^{(L)}}|\mathbf{0}\rangle$ is the (1,1)-coefficient of the matrix $\frac{1}{L}\,e^{it\mathcal{H}^{(L)}}\mathcal{N}^{(L)}\,e^{-it\mathcal{H}^{(L)}}$.

%

\end{document}